\documentclass[11pt,a4paper]{article}
\usepackage[utf8]{inputenc}
\usepackage[T1]{fontenc}
\usepackage{lmodern}
\usepackage{enumitem}

\usepackage[left=2.5cm,top=2.5cm,right=2.5cm,bottom=2.5cm]{geometry}
\usepackage{setspace}
\usepackage{amsmath,amssymb,amsfonts,amsthm,bm,latexsym}
\usepackage{nicefrac}

\usepackage{array}
\usepackage{booktabs}
\usepackage{longtable}
\usepackage{multirow}
\usepackage{float}
\usepackage{rotating}
\usepackage{pdflscape}
\usepackage{morefloats}
\usepackage{placeins}

\newcolumntype{C}[1]{>{\centering\arraybackslash}m{#1}}
\newcolumntype{P}[1]{>{\raggedright\arraybackslash}p{#1}}

\renewcommand{\arraystretch}{1.03}
\usepackage{graphicx}
\usepackage{svg}

\usepackage[authoryear]{natbib}
\usepackage[bottom]{footmisc}
\usepackage{xurl}
\usepackage{microtype}
\usepackage[none]{hyphenat}

\usepackage{color}
\usepackage{soul}
\usepackage{todonotes}
\usepackage{comment}

\usepackage{authblk}
\usepackage{multicol}
\usepackage{pifont}

\usepackage{xr-hyper}
\usepackage{hyperref}
\hypersetup{
    colorlinks=true,
    breaklinks=true,
    linkcolor=[rgb]{0,0.3,1},
    citecolor=[rgb]{0,0.3,1},
    urlcolor=[rgb]{0,0.3,1}
}

\newtheorem{theorem}{Theorem}

\newtheorem{claim}{Claim}
\newenvironment{cl}{\vskip.2cm\begin{claim}}{\end{claim}\vskip.2cm}

\newtheorem{lemma}{Lemma}

\newtheorem{corolla}{Corollary}

\newtheorem{defini}{Definition}
\newenvironment{defi}{\vskip.2cm\begin{defini}}{\end{defini}\vskip.2cm}

\newtheorem{proposi}{Proposition}

\newtheorem{prop}{Proposition}

\newtheorem{cla}{Claim}

\newtheorem{assump}{Assumption}

\newtheorem{demo}{Proof}

\newtheorem{remark}{Remark}

\newtheorem{example}{Example}

\newtheorem{notat}{Notation}

\def\cqfd {$\Box$}
\def\co{{\rm co}\hskip 1pt}

\def\E{{\mathcal E}}

\def\F{{\mathcal F}}

\def\H{{\mathcal H}}

\def\M{{\mathcal M}}

\def\N{{\mathbb{N}}}

\def\proj{{\rm proj}}
\def\proj0{{\rm proj}_{\co M_0}}

\def\R{{\mathbb{R}}}

\begin{document}
\title{Labour reallocation and production-network recomposition in the green transition}

\author[1,3,*]{Marco Due\~nas}
\author[2,3]{Antoine Mandel}
\affil[1]{\small Institute of Economics -- Sant'Anna School of Advanced Studies, Pisa, Italy}
\affil[2]{\small Centre d'Economie de la Sorbonne --  Paris School of Economics -- CNRS-Universit\'{e} Paris 1 Panth\'{e}on-Sorbonne, Paris, France}
\affil[3]{\small CLIMAFIN -- Climate Finance Alpha, Paris, France}
\affil[*]{Corresponding author email: marco.duenas@climafin.com}

\date{}

\maketitle
\vspace{-40pt}

\begin{abstract} 
\noindent 
A growing number of countries are adopting policies to reduce greenhouse-gas emissions, a process expected to reshape production systems and labour markets by altering relative prices and changing production and demand patterns across sectors. We develop an agent-based monetary input--output network model to analyse the medium-term effects of climate policies on production networks and sectoral labour allocation. The model features heterogeneous industries and households, incorporates greenhouse-gas emissions accounting, and is calibrated to a multi-regional input--output database covering the world economy. We simulate two alternative climate-policy scenarios over 2026--2035: one based on countries' nationally determined contributions and another targeting larger reductions in global emissions. The results reveal significant economy-wide reallocation: fossil-energy sectors contract, low-carbon energy sectors expand, and their positions within production networks change asymmetrically. Fossil energy loses relevance as a direct supplier and exerts less forward influence, while its remaining production becomes more dependent on intermediate inputs and upstream suppliers; low-carbon energy primarily strengthens its role as a supplier. Carbon pricing also generates sectoral labour gaps that, under mobility frictions, imply greater potential transitional unemployment. Because the model allows only limited technological change and adjustment within existing production links, the simulations provide a counterfactual assessment of how existing production structures respond to a relatively rapid climate transition, revealing the adjustment pressures that emerge when policy ambitions advance faster than the productive system’s capacity to adapt.
\end{abstract}

\bigskip
\noindent \textbf{Keywords:} Green transition; Production networks; Labour reallocation; Carbon pricing; Input–output analysis; Low-carbon economy.

\medskip
\noindent \textbf{JEL Codes:} Q52, D85, C63, C67
\medskip

\clearpage
\newpage
\section{Introduction}

The distributional consequences of climate policy are central to the social acceptability and the political sustainability of decarbonisation strategies. The transition to a low-carbon economy implies profound structural changes in production systems and labour markets, as policies such as carbon pricing alter relative prices across sectors and reshape production technologies, input--output linkages, and labour demand. While carbon pricing is designed to internalise environmental externalities, it also generates asymmetric adjustment costs across sectors, regions, and workers. These uneven impacts may amplify inequality and generate resistance to climate policies \citep{vona2021tasks, douenne2022yellow, dechezlepretre2025fighting}. Against this background, integrated-assessment research has focused primarily on aggregate macroeconomic outcomes and technological feasibility \citep[see][for recent surveys]{riahi2022mitigation, sarmiento2024comparing}, while devoting comparatively less attention to how climate-policy adjustment propagates jointly through production networks and sectoral labour markets.

This paper addresses this issue by developing an agent-based modelling framework to analyse structural adjustment under alternative climate-policy trajectories. We focus on three related questions: (i) how climate policy redistributes activity across sectors; (ii) how this reallocation changes the economic role of sectors within existing production networks; and (iii) how labour-market frictions affect the capacity of workers to move between declining and expanding activities. Rather than forecasting a unique transition path, the exercise provides a counterfactual assessment of the adjustment dynamics generated by alternative policy configurations and the capacity of the economic system to accommodate the resulting reallocation.
 
We use an agent-based input--output network model with endogenous production-network adjustment that generates out-of-equilibrium dynamics within a general-equilibrium framework by simulating agents that make input-allocation, production, and pricing decisions based on local interactions within their production network. We calibrate the model using the EXIOBASE multi-regional input--output database, covering 49 regions and 163 industries. Each country–industry pair is represented by one firm, and each firm employs intermediate inputs, capital and three skill-specific types of labour whose supply and wages are differentiated across sectors. At the country level, workers are represented by three skill-specific households that pool the income generated by their sectoral labour allocations, alongside one representative capital-owning household.

We build on the baseline model presented by \cite{mandel2020economic} and \cite{mandel2023disequilibrium} along two dimensions. First, we develop a framework to represent climate policy scenarios, introducing a detailed representation of greenhouse-gas (GHG) emissions that accounts for both process and combustion-related emissions. In addition, we introduce a detailed representation of carbon taxes and carbon permit markets that accounts for the dynamic evolution of market structure and sectoral coverage. Second, we develop a labour-market module that represents industry- and skill-specific labour supply and simulates inter-sectoral labour flows as a function of wage differentials and skill transferability across sectors. 

We simulate three focal climate policy scenarios in line with the Network for Greening the Financial System (NGFS) scenarios \cite[see:][]{ngfs2023scenarios, mandel2025new}: a business-as-usual scenario, a relatively lower-effort scenario consistent with nationally determined contributions under the Paris Agreement, and a faster transition scenario called ``Highway to Paris''. We focus on the European Union (EU), where the European Green Deal and Fit for 55 frame climate neutrality as an economy-wide and socially inclusive transformation, while the Clean Industrial Deal links decarbonisation to industrial competitiveness, skills, and quality employment \citep{europeancommission2019greendeal, europeancommission2021fitfor55, europeancommission2025cleanindustrial}. This policy setting makes the EU particularly relevant for examining how the effects of climate policy propagate through production networks and labour markets.

The results show that climate policy generates substantial structural reallocation whose magnitude and sectoral distribution depend on policy stringency and coverage. The most pronounced changes occur within the energy system: fossil-energy activities contract persistently, whereas low-carbon energy production expands strongly. Adjustment nevertheless extends well beyond a direct substitution between these two activities, affecting mining, manufacturing, transport, agriculture, services, and other sectors through changes in intermediate and final demand. Aggregate indicators alone therefore conceal the scale and heterogeneity of the underlying sectoral adjustment. 

The production-network response is primarily re-compositional. Because the set of feasible supplier--customer relationships is fixed in the model, climate policy operates through changes in the intensity and economic importance of existing production linkages rather than through unrestricted network rewiring. The response is asymmetric: fossil-energy sectors lose importance as direct suppliers and exert less forward influence, while fossil production becomes more dependent on intermediate inputs and on its upstream supply network; low-carbon energy activities primarily gain relevance as suppliers. Climate policy therefore redistributes not only the scale of sectoral activity but also sectors' functional position within the existing production structure.

Worker movements extend beyond a direct transfer from fossil to low-carbon energy, with large sectors such as services absorbing an important share of gross reallocation. More importantly, workers' ability to move across activities determines the persistence of the resulting labour-market mismatches. When skill transferability and mobility are limited, contracting sectors, particularly fossil energy and mining, accumulate labour surpluses while expanding activities, most notably low-carbon energy, simultaneously face vacancies. Faster mobility substantially reduces both types of labour gaps. Differences across skill groups emerge even though the simulations impose identical mobility speed parameters across skills, showing that sectoral exposure and the structure of reallocation opportunities themselves shape the distribution of adjustment. Therefore, policies that facilitate occupational and inter-sectoral mobility, such as retraining and support for worker transitions, can reduce the persistence of these mismatches by accelerating labour movement from contracting to expanding activities.

Overall, our paper contributes along three main directions. First, it extends climate-economy agent-based models by embedding country- and sector-specific carbon taxes and emissions caps into a globally calibrated multi-regional input--output system \citep{lamperti2020transitions, ciola2023matrix, amendola2024efficiency, ravaioli2025tackling, wieners2026policies}. Second, it adds to production-network research by allowing policy-driven price changes to shift real input use across existing supplier--customer links \citep{acemoglu2012network, aguilar2023networks}. Finally, it links these network changes to skill-specific labour markets and structured inter-sectoral mobility, both central to the adjustment costs of the green transition \citep{vona2021tasks, vandeplas2022labour}.

The remainder of the paper is organised as follows. Section~\ref{sec:literature} reviews the related literature. Section~\ref{sec:model} presents the modelling framework and simulation setup. Section~\ref{sec:results} discusses the main findings. Finally, Section~\ref{sec:concluding_remarks} concludes. Supplementary Material (SM) reports additional details on the data, model implementation, and auxiliary results.

\section{Related literature}
\label{sec:literature}

In our model setup, climate policy does not move the economy directly from one equilibrium to another; it sets the transition process that evolves through prices, production decisions, and labour relocation. Dynamic input--output systems, decentralised monetary exchange, inventories, and gradual price and quantity adjustment provide the classical foundations for analysing such paths \citep[see, for instance,][]{leontief1953structure, jorgenson1961stability, benassy1975neo, green1980disequilibrium}. 

Recent agent-based models (ABMs) operationalise this perspective by deriving macroeconomic dynamics from heterogeneous agents and their interactions. In climate applications, this permits micro-level damages, technological competition, and policy responses to generate non-linear and path-dependent aggregate outcomes \citep{balint2017complexity, lamperti2018faraway, lamperti2020transitions}. Energy-focused ABMs also show that shocks which look similar in aggregate can operate through different sectoral and distributional channels, and that technology policies with comparable environmental effects can have different macroeconomic and fiscal consequences \citep{ciola2023matrix, amendola2024efficiency}. For this paper, the key implication is that climate policy effects cannot be understood from aggregate outcomes alone, since they also depend on the mechanisms through which cost changes travel across sectors and reshape economic activity over time. This motivates closer attention to the international production structure through which policy-induced cost changes spread and alter sectoral activity.

Production-network models provide that sectoral perspective. Input--output linkages can transform idiosyncratic disturbances into aggregate fluctuations, depending on the position of the sectors initially affected \citep{acemoglu2012network}. Carbon pricing makes this mechanism especially salient: exposure reflects not only a sector's direct emissions but also the policy exposure of its suppliers and customers. In a general-equilibrium model calibrated to Spain, \citet{aguilar2023networks} find that an increase in the allowance price from approximately EUR~25 to EUR~100 per tonne of CO$_2$ causes a cumulative three-year gross domestic product (GDP) decline of 0.37\%, while sectoral value-added effects range from losses close to 4\% to virtually zero. The asymmetry depends on both direct emissions-trading coverage and network exposure, showing how moderate aggregate effects can mask pronounced sectoral adjustments.

\citet{ciarli2019structural} indicate that climate-economy model families differ substantially in whether they endogenise sectoral composition, industrial organisation, technology, employment, demand, and institutions. This distinction matters for our purposes because shocks propagate through observed input--output links differently from structural recomposition within those links. Related literature on directed technical change focuses on one margin of this recomposition: carbon prices and research support can redirect innovation toward clean technologies, while delayed intervention makes the transition slower and more costly \citep{acemoglu2012environment, acemoglu2016clean}. In this respect, our paper emphasises a complementary margin. The binary supplier--customer network is fixed, but firms change expenditure shares across feasible inputs when relative prices move; linkage intensities, centralities, and upstream or downstream positions consequently evolve. We describe this process as production-network recomposition or sectoral repositioning within production networks, not unrestricted topological rewiring. As this recomposition expands some activities and contracts others, its costs depend on firms' ability to substitute production inputs and on workers' ability to relocate across sectors.

Input--output estimates for a transition to renewable electricity show that changes in labour demand extend beyond energy production to upstream activities such as construction and machinery manufacturing \citep{cerny_etal_2022}. Evidence on green re-industrialisation in Europe similarly indicates that the employment and value-added effects of expanding green production are amplified by input--output linkages and existing industrial capabilities \citep{frattini2024green}. These findings shift attention from individual industry exposure to the reallocation of production across interconnected activities.

Regarding labour relocation, the literature indicates that such mobility is neither instantaneous nor uniform across industries. For Europe, \citet{vandeplas2022labour} argue that limited aggregate employment effects can coexist with substantial changes across sectors, occupations, firms, and regions, and that adjustment costs depend on the geographic and skill proximity of declining and expanding jobs. A task-based approach likewise treats occupational greenness and the capabilities required by green work as continuous rather than binary, making skill proximity informative about displaced workers' prospects \citep{vona2021tasks}. A relatedness perspective further shows that employment diversification is more likely when new jobs are connected to pre-existing regional capabilities, whether through task complementarities, skill similarity, or local synergies \citep{farinha2019relatedness}. Direct evidence from German social-security records provides a stronger basis for heterogeneous inter-industry mobility, where worker flows are concentrated in a sparse and stable set of industry pairs and excess flows are evidence of similarity in human-capital requirements \citep{neffke2017inter}. Using European industry data, \citet{marin2019skills} further show that higher energy prices have little effect on total employment but shift workforce composition towards technicians and away from manual workers.

These findings motivate a modelling framework in which wage signals induce gradual labour movements that are constrained by transferable capabilities; however, they do not directly identify the transition mechanism used in this paper. In our context, transferable capabilities are proxied by a similarity-based mechanism, a parsimonious representation of structured inter-industry moves. Labour studies typically observe mobility patterns or displacement after labour-demand shocks. Our framework complements them by generating the sectoral demand and wage changes jointly with production-network recomposition. This distinction matters because frictions in reallocating workers can generate substantial adjustment costs even when their aggregate employment effects appear limited.

Evidence on environmental regulation makes clear why net aggregates are insufficient measures of this adjustment. Using linked worker--firm data around the 1990 Clean Air Act Amendments, \citet{walker2013transitional} finds that the average worker initially employed at a newly regulated plant loses about 20\% of pre-regulation annual earnings in present-value terms. The losses arise primarily through job separation, non-employment, and lower subsequent earnings, even though their aggregate value is small relative to the estimated health benefits of the regulation. In a two-sector general-equilibrium model with search frictions, \citet{hafstead2018unemployment} show the corresponding aggregate mechanism: an emissions tax produces large employment losses in the polluting sector and offsetting gains in the non-polluting sector, leaving a much smaller net employment effect. 

However, across the wider economy, carbon-pricing evidence reinforces the idea that a moderate aggregate cost may still produce a substantial and costly process of worker reallocation. British Columbia's revenue-neutral carbon tax generated a small positive aggregate employment effect, but employment fell in emissions-intensive, trade-exposed industries and rose in cleaner services \citep{yamazaki2017jobs}. Worker-level analysis also finds disproportionate adverse effects for less-educated workers, although the estimates may partly reflect an unobserved provincial labour-market shock \citep{yip2018labour}. 

We bring these strands together in a globally calibrated agent-based framework. Our contribution is to connect climate-policy dynamics and production-network adjustment to frictional labour reallocation, thereby explaining how moderate aggregate responses can coexist with pronounced sectoral and structural change, and how the cost of that change depends on the frictions governing worker mobility.

\section{Economic model}
\label{sec:model}

\subsection{General equilibrium framework}

The economy consists of $C$ countries, each comprising the same $L$ industries. We calibrate the model to the world economy using the EXIOBASE database \citep{merciai2018methodology}, which provides input--output relationships among $L=163$ industries across $C=49$ regions. The data set includes the twenty-seven EU member countries, seventeen major economies and five aggregated rest of the world regions. Within our model, each industry--country pair is represented by a single firm supplying a differentiated variety. Each of the $C$ countries also has four representative households: low-, medium-, and high-skill workers, each deriving income from supplying the corresponding skill-specific labour services; and one capitalist who derives income from the rent of capital services.

The set of goods $\M$ in the economy corresponds to the output of the different types of agents: the three types of skilled labour, the capital good, and the output of each type of industry, all differentiated by country. Thus, there are $M=| \M |=C(L+4)=8183$ such goods and therefore the same number of agents. We shall denote by $\F \subset \M$ the set of industries (and the corresponding outputs) and by $\H=\H_L\cup \H_K$ the set of households, comprising workers and capitalists. 

Each firm $i$ produces its differentiated good using capital provided by the domestic capitalist, the different types of labour provided by domestic workers, and a combination of domestic and international inputs. More precisely, the production technology of each firm $i$ is described by a nested constant elasticity of substitution (CES) production function of the form:
\begin{equation} \label{prodfun}
f_{i}(x_i) = \Lambda_i \phi_i(x_i)
\end{equation}
with, 
\begin{equation}
\phi_i(x_i) =
\left[
  \sum_{s \in S_i}
  \left(
    \sum_{j \in s} \beta_{ji} x_{ji}^{\rho_s}
  \right)^{\theta / \rho_s}
\right]^{1/\theta}.
\end{equation}
where $\phi_i$ denotes the underlying physical production technology, $\Lambda_i$ the productivity level,\footnote{In the baseline equilibrium, productivity is normalized to $\Lambda_i=1$ for all firms.} $x_{i}=(x_{ji})_{j \in \M}$ is the vector of inputs of the firm $i$, and $S_i$ is the corresponding nest partition of inputs. For firms, we partition the set of inputs into $S=8$ categories: ``value added'' (corresponding to labour and capital inputs, i.e. $\H$); agriculture and forestry; energy services; manufacturing; digital services; services; fossil inputs; and mining. See the complete classification in Table~\ref{SM-tab:S_industries}.

Each type of worker $h \in \H_L$ provides a fixed amount of work $e_h \in \R_+$ and each capitalist $h \in \H_K$ provides a fixed amount of capital services $e_{h} \in \R_+$ every period. Households also have nested CES preferences of the form
\begin{equation} \label{utfun}
u_{h}(x_h) = 
\left[
  \sum_{s \in S'}
  \left(
    \sum_{j \in s} \beta'_{jh} x_{jh}^{\rho'_s}
  \right)^{\theta' / \rho'_s}
\right]^{1/\theta'}.
\end{equation}
where $(x_h)_{h \in \H}$ is the consumption vector of the household $h$, and $S'$ is the household nest partition, which is assumed to be identical across households. For households, we partition the set of goods into $S=8$ categories: agriculture and food; energy services; manufacturing; digital services; transport services; services; waste; and mining. See the complete classification in Table~\ref{SM-tab:S_industries}. 

We denote the economic system introduced above as $\E(\M,\beta,\rho,\theta).$  In this setting, a general equilibrium is standardly defined as follows.

\begin{defi} A general equilibrium of the economy $\mathcal{E}(\mathcal{M},\beta,\rho,\theta)$ is a 
collection of prices $p^* \in \R^M_+$, production levels 
$y^* \in \R^M_+$ and commodity flows $(x^*_{j,i})_{j,i\in \mathcal{M}} \in \R^{M\times M}_+$ such that:
\begin{enumerate}

\item Consumers maximise their utility under their budget constraint. That is for all $h \in \mathcal{H},$ $x^*_h \in \R^M_+$ is a solution to 
$$\left\{\begin{array}{c} \max u_h(x_h) \\  \\
\text{s.t }   p^* \cdot x_{h} \leq  p^*_h e_h \end{array}\right.$$

\item Firms maximise profits. That is for all $i \in \mathcal{F},$ 
$(y^*_i,x^*_i) \in \R_+ \times \R^M_+ $ is a solution of 
$$\left\{\begin{array}{c} \max  p^*_i y_i- p^* \cdot x_{i}  \\  \\
\text{s.t }  f_i(x_i)\geq y_i \end{array}\right.$$

\item Markets clear. That is, for every $j \in M,$ one has 
$$y^*_j = \sum_{i=1}^M x^*_{j,i}.$$
Furthermore, for all $j \in \mathcal{H},$ $y^*_j=e_j$ to account for the inelastic supply of labour and capital services in the economy.
\end{enumerate}
\label{monequi}
\end{defi}

\begin{remark}
 Note that, given constant returns in production technologies, there are no profits to be distributed at equilibrium. Namely, one has for all $j \in \mathcal{F}$
 $$p^*_j y^*_{j} =p^* \cdot x^*_{j} $$
The non-profit condition implies, in particular, that households' income is entirely determined by their supply of labour or capital services.
\end{remark}

As highlighted in the recent macroeconomic literature \cite[see, e.g.,][]{baqaee2019macroeconomic,carvalho2019production}, the general equilibrium of this economy defines nominal input--output flows between each pair of agents. These flows provide the equilibrium expenditure-share representation used to initialise the model. The structure of this network can be captured by a column-stochastic matrix $A^*=(a^*_{i,j})_{i,j \in \mathcal{M}} \in \R^{M \times M}_+$ whose coefficient $a^*_{i,j}$ captures the share of $j$'s equilibrium spending directed towards $i,$ that is  
\begin{equation} 
a^*_{i,j}=\dfrac{p^*_i x^*_{i,j}}{p^* \cdot x^*_{\cdot,j}}.
\end{equation} 
In particular if $j \in \mathcal {F}$ and $i \in \mathcal {F},$ $a^*_{i,j}$ represents the amount of intermediate consumption spent on good $i$ per unit of revenue of firm $j$ while for $i \in \mathcal {H}$, $a^*_{i,j}$ represents the value added received by household $i$ per unit of revenue of firm $j.$

\subsection*{Agent-based dynamics}

Building on the conceptual framework introduced in \citep{axtell2005complexity, gintis2007dynamics} and the model developed in \citep{gualdi2016emergence, mandel2020economic}, we ``agentify" this general-equilibrium model in order to analyse out-of-equilibrium dynamics induced by climate policy shocks and their distributional consequences. Formally, we consider discrete periods of time indexed by $t \in \N.$ In the baseline version of the model, the state of the system in period $t$ is determined by the following variables for each agent $i \in \M:$
\begin{itemize}
\item Monetary balance $m_i^t,$ corresponding to the consumption budget for households and the working capital for firms.
\item Budget shares $a^t_i=(a_{j,i}^t)_{j\in\M} \in \R_+^M$ corresponding to the proportion of expenses of agent $i$ directed towards input $j$. 
\item The price $p_i^t$ at which the output is being sold (for workers and capitalists, these correspond to wage and capital rental rates, respectively).
\item The stock of output $q_i^t \in \R_+$ (constant and equal to $e_i$ for household agents $i \in \H$).  
\end{itemize}

\noindent 
The state of each agent then evolves dynamically through the following sequence of interactions \cite[see][for details]{mandel2020economic,mandel2021monetary}:
\begin{enumerate}
\item  Each agent $i \in \M$ receives the nominal demand $\sum_{j \in \M} a^t_{i,j} m^t_j.$ 

\item Agents adjust their prices frictionally towards their market-clearing values according to: 
\begin{equation}
p_i^t=\tau_p \overline{p}_i^t+ 
(1-\tau_p) p^{t-1}_i \label{pricefric}\;;
\end{equation}
where $\tau_p \in[0,1]$ is a parameter measuring the speed of price adjustment and $\overline{p}_i^t$ is the market-clearing price defined as follows. Given the nominal demand $\sum_{j \in \M} a^t_{i,j} 
m^t_j$ and the output stock $q_i^t,$ the market clearing price for agent $i$ is: 
\begin{equation}\label{price}
\overline{p}_i^t=\dfrac{\sum_{j \in 
\M} a^t_{i,j} m^t_j}{\nicefrac{q_i^t}{\chi}} \;. 
\end{equation} 
where $\chi$ is the target inventory-to-output ratio.

\item Whenever $\tau_p<1,$ markets fail to clear (except if the system is at a stationary equilibrium). In case of excess demand, we assume that buyers are rationed proportionally to their demand, i.e. the quantity of good $j$ purchased by agent $i$ is given by:
\begin{equation} 
x^t_{j,i} =\min\left(1, \frac{q^t_j/\chi}{\sum_{k \in \M} a^t_{j,k}  m^t_k/p_j^t}\right) \frac{a^t_{j,i}m^t_i}{p_j^t}  \;.
\label{input} 
\end{equation}

\item Firms update their inventory by producing new output and accounting for sales:
\begin{equation}
q^{t+1}_i=q^t_i-\sum_{k \in \M} x_{i,k}^t + f_i(x_i^t) 
\label{prod}
\end{equation}

\item Agents update their monetary balances on the basis of their purchase of inputs and their sales of output\footnote{Equation~\eqref{evmoney} can be interpreted as assuming that firms have myopic expectations about their nominal demand (i.e. firms assume they will face the same nominal demand next period) and target a balanced budget (net of labour and capital costs paid to households).}, i.e.
\begin{equation}
m^{t+1}_i=m^{t}_i  + \sum_{k' \in \M} p_i^t x_{i,k'}^t - \sum_{k \in \M} p_k^t x_{k,i}^t  
\label{evmoney} 
\end{equation}

\item Agents adjust frictionally their input/consumption combinations towards the cost-minimising value: 
\begin{equation} 
a_{i}^{t+1}=\tau_{w} \overline{a}_{i}^{t} +(1-\tau_{w}) a_{i}^t \;;
\label{optweightfric}
\end{equation}
where $\tau_{w} \in[0,1]$ measures the speed of technological/preferences adjustment and $\overline{a}_{i}^{t} \in \R^M$ denotes the vector of optimal expenditure shares for agent $i$. For firms, these shares maximise production, while for households they maximise utility, given prevailing prices. Formally, the optimal expenditure shares are defined by\footnote{The fraction in Eq.~\eqref{optinput} is taken coordinate wise.}:
\begin{equation} 
\overline{a}_{i}^{t}:= \operatorname*{\text{argmax}}_{\{a_i \mid \sum_{j\in\M} a_{j,i}=1\}} g_i\left(\frac{a_{i}}{p^t}\right) \:,
\label{optinput}
\end{equation}
where $g_i=f_i$ if $i\in\F$ and $g_i=u_i$ if $i\in\H$ for households.

\item Firms accumulate production experience over time. We capture learning-by-doing through a productivity parameter $\Lambda_i^t$ multiplying the production function. Let $Q_i^t$ denote the cumulative production experience of firm $i$ up to period $t$, learning-by-doing is introduced through
\begin{equation}\label{eq:learning}
\Lambda_i^{t+1} = \left(\frac{Q_i^{t}}{Q_i^{0}} \right)^{\lambda_i},
\end{equation}
where $\lambda_i\geq 0$ is the learning parameter of sector $i$. In sectors without learning-by-doing, $\lambda_i=0$, implying $\Lambda_i^{t+1}=\Lambda_i^t$.
\end{enumerate}

The agent-based dynamics have been extensively analysed in previous contributions \cite[see, e.g.,][for details]{gualdi2016emergence,mandel2021monetary}. If the price adjustment is fast, the economy converges to the underlying general equilibrium. If price adjustment is slow, inventories carry the burden of adjustment and the economy is characterised by a synchronised state with large ``cyclical volatility''. In empirical applications \citep[see, e.g.,][]{mandel2020economic, mandel2023disequilibrium}, considering out-of-equilibrium dynamics allows us to provide a precise description of shock propagation mechanisms and of the resulting macroeconomic response.

\subsection{Labour market model}

In the baseline dynamics introduced above, we abstract from frictions in the reallocation of workers across sectors. A more precise representation of the labour market is required to analyse the distributional and structural impacts of climate and energy policies, which entail major structural changes and require substantial reallocation of the workforce across industries \citep{dechezlepretre2023}. Accordingly, we extend our baseline model with a detailed representation of the labour market, in which wages and labour supply are set separately in each industry, and inter-sectoral labour flows are frictional and constrained by the transferability of skills.  

More formally, we represent each worker type $h \in \H$ as a continuous mass split across the different sectors/industries in the country. Namely, for a given country, we denote by $\ell^t_{h,i} \in \R_+$ the labour supplied by workers of type $h$ to industry $i$ in period $t,$ which is such that $\sum_{i \in \F} \ell^t_{h,i}= e_h.$ We further denote by $p^t_{h,i} \in \R_+$ the wage for workers of type $h$ in industry $i$. The first extension of the baseline model we consider is that wages are set independently in each sector, taking into account the industry-specific supply and demand of labour. Namely, step 2 of the baseline model is extended as follows. 

\begin{enumerate}[label=2'.]
\item The wage updating mechanism for industry $i$ and worker type $h$ is given by:  
\begin{equation}\label{labourpricefric}
p^t_{h,i}=\tau_p \overline{p}^t_{h,i} + (1-\tau_p) p^{t-1}_{h,i} 
\end{equation}
where $\overline{p}^t_{h,i}:= \nicefrac{a^t_{h,i} m^t_i}{\ell^t_{h,i}}$ is the clearing price on the labour market for worker-type $h$ in sector $i$.
\end{enumerate}

The second major extension we consider is that the supply of labour across sectors evolves dynamically. This evolution is driven by the magnitude of the wage differential between sectors and constrained by the similarity of skills/occupations across industries. More specifically, we consider there to be excess supply of labour of type $h$ in industry $i$ if the wage $p^t_{h,i}$ is lower than the average wage  $\tilde{p}^t_h:= \nicefrac{ \sum_{i \in \F} p^t_{h,i} \ell^t_{h,i}}{e_h}.$ Conversely, there is excess demand for labour if the industry wage is greater than the average wage. In order to quantify the magnitude of the excess demand/supply of labour, we benchmark it against the labour supply that would make the wage bill per unit of labour equal across sectors (by making the ratio between total wage bill and labour supply constant), that is:
\begin{equation} 
\overline{\ell}^t_{h,i}:= \ell^t_{h,i} \dfrac{p^t_{h,i}}{\tilde{p}^t_h}
\end{equation} 
Accordingly, the overcapacity (or conversely, minus the number of vacancies) in a sector is given by  
 \begin{equation} \label{eq:labour_gap}
\Delta \ell^t_{h,i}:= (\ell^t_{h,i}-\overline{\ell}^t_{h,i}) \;. 
\end{equation}
Therefore, $\Delta \ell^t_{h,i}$ measures the gap between the available labour and the allocation implied by equalising remuneration per hour across sectors within each country–skill group.

We measure the capacity of workers to transition from sector $i$ to sector $j$ through a sectoral similarity index $s_{i,j}$, defined as the cosine similarity between sectoral input vectors:
\begin{equation}
s_{i,j} = \frac{z^*_i \cdot z^*_j}{\|z^*_i\| \, \|z^*_j\|}
\end{equation}
where $z^*_i$ and $z^*_j$ denote the columns corresponding to sectors $i$ and $j$ in the baseline real input-coefficient matrix $\mathbf Z^*=(a^*_{i,j})_{i,j\in\mathcal F}$, obtained as the firm-to-firm submatrix of $A^*$. Because all prices are normalised to one in the reference equilibrium year, these coefficients can be interpreted in real terms. The numerator is the Euclidean inner product of the two vectors and $\|\cdot\|$ denotes the Euclidean norm. Hence, higher values of $s_{i,j}$ indicate greater similarity in input structures across sectors and are therefore associated with a higher likelihood of labour mobility, as workers can more easily transfer their skills across technologically related activities.

Sector similarity generates a relatively dense network of potential labour transitions. To avoid unrealistically high levels of labour mobility, we sparsify the network by retaining only links above a common similarity threshold. The threshold is chosen so that the minimum density across country-level mobility networks is approximately 10\%, in line with estimates based on observed worker inter-industry transitions reported by \cite{neffke2017inter}. This results in a threshold value of $\bar{s}=0.25$.

We then extend the baseline model with the following dynamics of labour supply.
\begin{enumerate}[label=8.]
\item The supply of labour by worker type $h$ in sector $i$ is given by
\begin{equation}\label{eq:labour_supply}
 \ell^{t+1}_{h,i}= \ell^{t}_{h,i} + \tau_h
 \left\{
\begin{array}{ll} 
  \sum_{\{j \mid \Delta \ell^t_{h,j} >0 \}} \Psi^t_{j,i},
  & \text{if } \Delta \ell^t_{h,i} < 0, \\[1ex]
 (-)\sum_{\{j \mid \Delta \ell^t_{h,j} <0 \}} \Psi^t_{i,j},
  & \text{if } \Delta \ell^t_{h,i} > 0 .
 \end{array}\right.
\end{equation}
where $\tau_h$ is a parameter controlling the speed of adjustment of labour supply, while the magnitude of adjustment $\Psi^t_{i,j}$ is determined as follows:

Let $i$ be a sector with labour overcapacity, so that $\Delta \ell^t_{h,i}>0$. The potential flow of workers of type $h$ from sector $i$ to sector $j$ is given by
\[
\psi^t_{i,j} = \Delta \ell^t_{h,i} 
\frac{\max(p^t_{h,j}-p^t_{h,i},0)s_{i,j}}
{\sum_k \max(p^t_{h,k}-p^t_{h,i},0)s_{i,k}},
\]
where, for the sake of simplicity, we removed the skill subindex $h$. We then use the expected total inflow $\sum_k \psi^t_{k,j}$ into industry $j$ to scale these flows so that they do not exceed the number of vacancies in sector $j$, where $\Delta \ell^t_{h,j}<0$. The effective flow of workers from sector $i$ to sector $j$ is therefore given by
\begin{equation}
\Psi^t_{i,j} = \psi^t_{i,j} \min
\left( \frac{|\Delta\ell^t_{h,j}|} {\sum_k \psi^t_{k,j}}, 1 \right).
\end{equation}

\end{enumerate}
Hence, the scaling factor ensures that the total effective inflow into sector $j$ does not exceed the number of vacancies ($|\Delta\ell^t_{h,j}|$). The magnitude of labour flows from sector $i$ to sector $j$ depends on vacancies in sector $j$, the surplus of employees in sector $i$ ($\Delta \ell^t_{h,i}$), the wage differential $p^t_{h,j}-p^t_{h,i}$ and the similarity between sector $i$ and $j,$ $s_{i,j}$. It is straightforward to verify that labour supply dynamics are at a rest point when the wage is constant across sectors. Otherwise, labour reallocation tends to reduce wage differentials by decreasing (increasing) labour supply in sectors where wages are below (above) the average wage for the corresponding worker type. Thus, given a change in wages, the reallocation of workers depends on the parameter $\tau_h$ and on the transferability of skills, which is measured by the similarity matrix $(s_{i,j})_{i,j \in \mathcal{F}}.$

\subsection{Climate policy simulation setup}
\label{sec:green_calibration}

In the following, we focus on the out-of-equilibrium dynamics induced by the introduction of climate policy objectives in the economy. From this perspective, we first integrate a representation of direct firm-level GHG emissions into the model. 

On the basis of EXIOBASE data, which provides emissions related to different processes for each industry, we distinguish two sources of emissions: (i) \emph{output-related} emissions arising from the production process other than fuel combustion and (ii) \emph{input-related} emissions associated with the combustion of fossil fuels used as input (see Table~\ref{SM-tab:S_policy_industries} for the fossil input classification). More specifically, we define the emission intensity of each firm $i$ through two rates: $\iota^{\text{out}}_i,$ which gives the amount of emissions per unit of output produced and $\iota^{\text{in}}_i,$ which gives the amount of emissions per unit of fossil input used. Accordingly, total emissions of firm $i$ given the input $x_i$ is
\begin{equation}\label{eq:ghge_intesities}
g_i(x^t_i)= \iota^{\text{out}}_i \phi_i(x^t_i) + \iota^{\text{in}}_i \sum_{j \in \mathcal{F}_f } x^t_{j,i} \;;
\end{equation}
where $\mathcal{F}_f$ denotes the set of fossil fuel inputs.\footnote{In EXIOBASE, combustion-related emissions are reported at the sectoral level and are not consistently disaggregated by fossil fuel type (e.g., gas, oil, coal). As a result, we do not assign input-specific emission factors. Instead, we interpret $\iota^{\text{in}}_i$ as an average emission intensity per unit of fossil input, computed as the ratio between total combustion emissions and total use of inputs with combustion potential in sector $i$.}

We consider that firms can be subject to two different climate policy measures: a carbon tax or a cap-and-trade system. These affect firms' behaviour as follows.
\begin{itemize}
\item If a firm $i$ is subject to a carbon tax, whose rate in period $t$ is $\tau^t_c,$ it ought to pay, when using a vector of inputs $x_i^t \in \R^M_+$ an amount $\tau^t_c g_i(x^t_i)$ on top of its input costs. The evolution of its monetary balance, given by Eq.~\eqref{evmoney}, is accordingly modified. Furthermore, it accounts for the carbon tax in the updating of its input mix so that the vector of optimal input weights (see Eq.~\eqref{optinput}) is given by: 
\begin{equation} 
\overline{a}_{i}^{t}:= \operatorname*{argmax}_{\{a_i \mid \sum_{j \in \M} a_{j,i}+ \tau^t_c g_i(\nicefrac{a_{i}}{p^t})=1\}} 
f_i\left(\frac{a_{i}}{p^t}\right) \;. 
\label{optinput2}
\end{equation}
Tax revenues are redistributed to domestic households in proportion to their budget shares.
   
\item If a firm is subject to a cap-and-trade system $m$ covering the subset of firms $\mathcal{F}_m^t$ and faces a carbon price $p_{c,m}^t$, its individual behaviour is modified as if it were facing a carbon tax of the same magnitude. More precisely, its optimisation problem is identical to that under a carbon tax, replacing the tax rate with the endogenous carbon price.

The carbon price, similarly to that of other goods, is determined by equating the total nominal demand for allowances by regulated firms to the supply of allowances available in the market. We assume rapid adjustment of the emissions trading market (i.e. $\tau_p=1$ using the notation of Eq.~\eqref{pricefric}), so that the carbon price is given by
\begin{equation}
p_{c,m}^t =
\frac{ \sum_{i\in\mathcal{F}_m^t}
m_i^t g_i\left(\frac{a_i}{p^t}\right)}
{E_m^t},
\label{eq:ets_price}
\end{equation}
where $E_m^t$ denotes the quantity of emission allowances available in market $m$ during period $t$. Finally, revenues generated by the sale of emission allowances are redistributed to domestic households in proportion to their budget shares.
\end{itemize}

We define a climate policy scenario by specifying: (i) the set of firms subject to a carbon tax or to a cap-and-trade, (ii) the temporal profile of carbon tax rates, and (iii) the structure of the cap-and-trade, including its partition into submarkets and the temporal profile of emission allowances.

More formally, we denote by $\mathcal{F}_{\tau}\subset\mathcal{F}$ the set of firms subject to a carbon tax and by $\mathcal{F}_{m}\subset\mathcal{F}$ the set of firms subject to a cap-and-trade, either of which may be empty. For firms subject to a carbon tax, tax rates may vary across firms and over time and are given by $(\tau^t_{c,i})_{t\in\mathbb{N},\, i\in\mathcal{F}_{\tau}}$. For the cap-and-trade system, we assume that, at time $t$, regulated firms are partitioned into submarkets $(\mathcal{F}^t_{m,\mu})_{\mu\in I}$, and that the supply of allowances in submarket $\mu$ during period $t$ is given by $E^t_{m,\mu}$. Overall, a climate policy scenario specifies the temporal evolution of the constraints faced by firms in relation to their GHG emissions.

Such climate policies affect economic dynamics through several channels: (i) carbon taxes or carbon prices increase the relative cost of fossil-fuel-intensive inputs, inducing substitution towards cleaner alternatives, with a magnitude that depends on the elasticity of substitution; (ii) relative commodity prices adjust in response to differences in embodied emissions; and (iii) changes in relative prices lead to a reallocation of both final and intermediate demand across commodities. These adjustments feed back into the production structure, generating changes in the relative size of firms and industries as well as in the structure of input--output linkages. 

\subsection*{Calibration}

The model is calibrated in two stages. First, the economy is initialised in a stationary monetary equilibrium that exactly reproduces observed worldwide input--output relations in the base year 2019 (see Section~\ref{SM-sec:S_model_calibration}). Second, climate-policy parameters are chosen to approximate the effective stringency of decarbonisation policies observed between 2019 and 2025, while allowing forward-looking scenario simulations up to 2035.

The initial state of the model corresponds to a general equilibrium of the monetary input--output economy as described in Section~\ref{sec:model}, calibrated to EXIOBASE version~3.9.6 for the year 2019. For given values of the production-function elasticities and nesting structure, all share and efficiency parameters are calibrated so that the observed sectoral outputs, intermediate input flows, factor payments, and household expenditures are exactly reproduced. Prices are normalised to unity in the base year, and the resulting equilibrium features zero excess profits and balanced monetary flows across all agents.

EXIOBASE environmental satellite accounts are then used to calibrate emissions. We compute country-sector-specific emissions intensities of total GHG emissions (CO$_2$, CH$_4$, and N$_2$O), distinguishing between output-related (process) emissions and input-related emissions associated with the use of fossil fuels. These intensities are fixed at their 2019 levels and used to compute simulated emissions associated with output and input flows. This ensures that baseline emissions in the model match observed emissions in the base year. 

\begin{figure}[h!]
\centering
\includegraphics[width=0.99\linewidth]{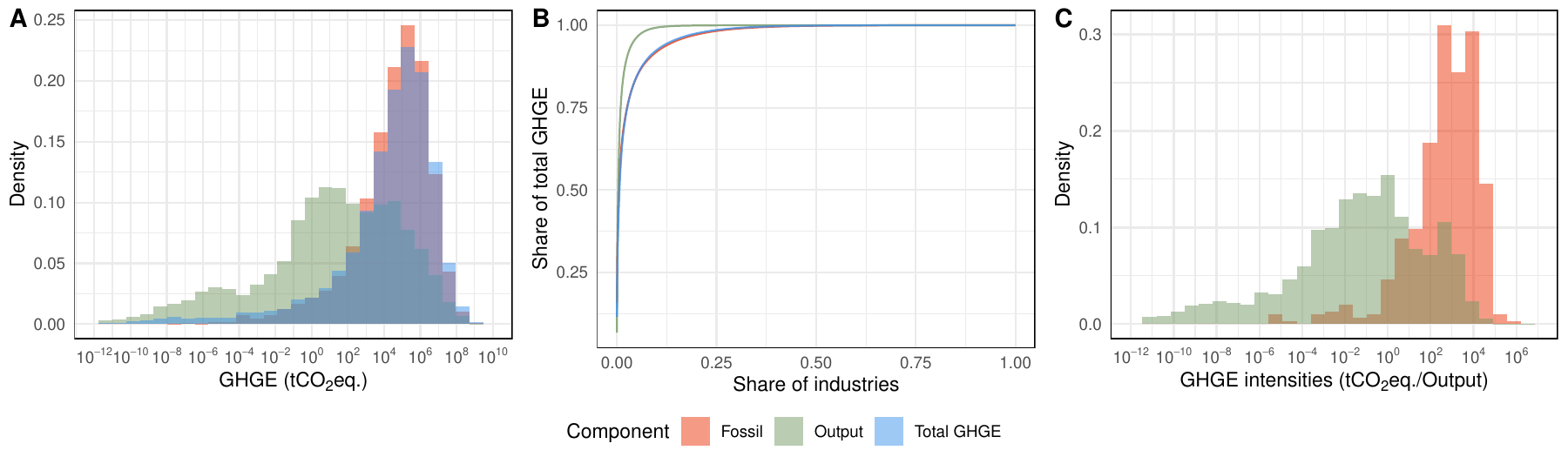}
\caption{Distribution of GHG emissions across country-industries in the 2019 calibration. Panel A reports total, fossil-input-related, and output-related emissions; Panel B reports their concentration across country-industries; and Panel C reports emission intensities.}
\label{fig:ghge_hist}
\end{figure}

Figure~\ref{fig:ghge_hist} presents the distribution of GHG emissions across country-industries related to output and fossil inputs and their corresponding intensities as presented in Eq.~\eqref{eq:ghge_intesities}. Panel A shows that total emissions and their fossil- and output-related components are highly right-skewed, indicating substantial heterogeneity in absolute emission levels. To complement this, Panel~B exposes a marked concentration pattern, in which a small share of country-industries accounts for a disproportionately large fraction of total emissions, as evidenced by the steep cumulative curve. Panel~C further indicates that emission intensities are also quite heterogeneous, with a heavy upper tail, indicating how different production structures contribute significantly to overall emission disparities.

\begin{figure}[h!]
\centering
\includegraphics[width=0.9\linewidth]{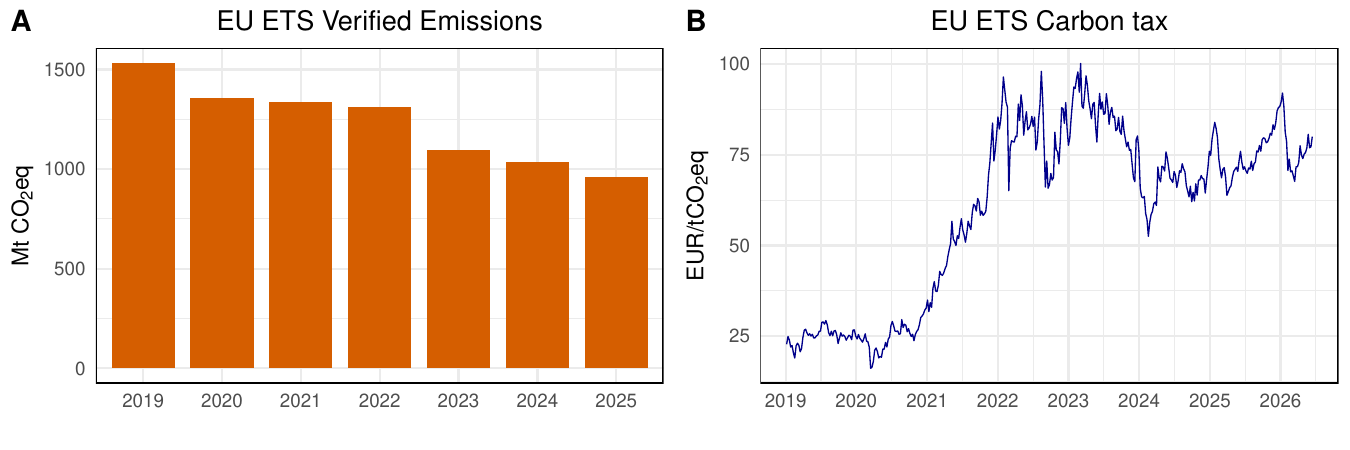}
\caption{Historical EU ETS emissions and carbon-price series used for model calibration. Panel A reports verified EU ETS emissions obtained from the European Environment Agency's EU ETS Data Viewer, based on data from the European Commission's Union Registry \citep{eea_euets_data}. Panel B reports the EUA yearly futures price traded on the European Energy Exchange, obtained from Investing.com \citep{investing_eua_data}.}
\label{fig:eu_ets_stat}
\end{figure}

Climate-policy parameters are calibrated to reflect the degree of decarbonisation observed between 2019 and 2025. Given the heterogeneity of climate policies across countries, we do not attempt to replicate individual regulatory instruments in full institutional detail. Instead, we identify country--industry carbon markets and assign them stylised carbon-pricing regimes---either carbon taxes or an emission cap---whose coverage and stringency are chosen to match observed emissions trajectories and, where available, observed carbon prices (see Section~\ref{SM-sec:S_policy} for details). Specifically, for the calibration, we use data from EU Emissions Trading System (EU ETS) emissions reports, which show a decline in emissions of around 35\% between 2019 and 2025 (see Panel A in Figure~\ref {fig:eu_ets_stat}). Implementing the corresponding emission pathway in the model leads to a carbon price ranging between 55 and 70 EUR, depending on elasticity values. For comparison, in 2025, the minimum observed price was EUR~63 at the end of March, and the maximum was EUR~87 at the end of December. Since 2022, historical prices have exhibited significant fluctuations, yet have broadly remained around EUR~70 (see Panel B in Figure~\ref {fig:eu_ets_stat}).

\subsection*{Climate policy scenarios}

During the 2019--2025 calibration period, the historical pattern of taxes in terms of policy coverage, country- and sector-specific regimes, and emission allowances is as described in Section~\ref{SM-sec:S_policy}. For the 2026--2035 period, we consider one business-as-usual and two scenarios corresponding to different levels of climate-policy stringency. These two scenarios extend coverage to all industries assigned to a climate-policy sector, excluding those classified as ``Other'' (see Table~\ref{SM-tab:S_policy_industries}). The EU is treated separately: its broad policy coverage is defined as the union of industries covered by the EU ETS and the Effort Sharing Regulation (EU ESR). Industries belonging to neither group, including mining, therefore remain outside direct climate-policy coverage in the EU. Therefore, for the whole time window, we consider 
\begin{itemize}
    \item A business-as-usual (BAU) scenario in which climate policy targets remain at their 2025 level throughout the 2026-2035 period.   
    \item A Nationally Determined Contributions (NDC) scenario, in which countries implement their commitment under the Paris Agreement, implying a gradual tightening of emissions constraints beyond 2025 until 2035. This leads to a 50\% reduction in GHG emissions in Europe (59\% in the EU ETS and EU ESR sectors) and a 30\% reduction globally by 2035 compared to 2019.
    \item A Highway to Paris (HWTP) scenario in which all countries implement a sector-wide carbon tax target of EUR 200/tCO$_2$, similar to the NGFS--HWTP scenario \cite[see][]{ngfs2023scenarios, mandel2025new}. This scenario is more stringent than the NDC and is assumed to be consistent with a net-zero emission objective in 2050, leading to a 55\% reduction in GHG emissions in Europe (62\% in the EU ETS and 50\% in the EU ESR sectors) and a 47\% reduction globally by 2035 compared to 2019.
\end{itemize}

For EU ESR sectors, differences in policy design imply that emissions reductions may be larger under the NDC scenario than under HWTP. In our calibration, the carbon price in EU ETS sectors reaches approximately EUR 129/tCO$_2$ under the NDC scenario. In contrast, the implicit carbon price required in EU ESR sectors reaches approximately EUR 311/tCO$_2$ under the NDC scenario. Consequently, for EU ESR sectors, although HWTP represents a more ambitious global decarbonisation pathway, the NDC scenario can generate larger emissions reductions because it imposes stronger constraints.

Finally, in all scenarios, carbon policy revenues are allocated to households as lump-sum transfers. Specifically, revenues are redistributed at the country level in proportion to households' money and further allocated across skill groups according to their income shares. Overall, climate policy interacts with economic dynamics through three main channels. First, carbon pricing introduces a price wedge that raises the relative cost of emission-intensive inputs and outputs. Second, the emission budget translates into effective quantity restrictions on production. Third, recycling carbon revenues to households creates an income-transfer channel that feeds back into consumption demand and distributional outcomes. 

\subsection*{Simulation setup}
\label{subsec:simulation_setup}

All simulations start from the calibrated 2019 baseline equilibrium, a fixed point of the baseline model, so subsequent dynamics are driven by climate policies and endogenous adjustment mechanisms.

The resulting out-of-equilibrium dynamics are governed by (i) the elasticities of substitution between and within nests for firms (see Eq.~\eqref{prodfun}) and consumers (see Eq.~\eqref{utfun}) that govern the extent to which firms and consumers adjust intermediate and final consumption following policy-induced price changes, (ii) the speed $\tau_p$ of price and $\tau_w$ of technology adjustments, and, more fundamentally for our analysis, (iii) the speed of adjustment of labour supply across industries $\tau_h$. 

The benchmark and range of parameter values considered in the simulation are reported in Table~\ref{tb:sim_parameters}. 
\begin{table}[h]
\centering
\caption{Model benchmark calibration parameters and sensitivity ranges (in parentheses).}
\label{tb:sim_parameters}
\resizebox{0.9\textwidth}{!}{%
\begin{tabular}{l >{\raggedright\arraybackslash}p{12cm} >{\raggedright\arraybackslash}p{6cm}}
\toprule
Parameter & Description & Benchmark / calibration range \\
\midrule
\multicolumn{3}{l}{\textit{Time structure}} \\
$\Delta t$ & Length of one model period & weekly: $52$ periods per year \\
$T$ & Total simulation horizon & 2019–2040; policy trajectories defined through 2035 plus 5 years to confirm stability \\
\midrule
\multicolumn{3}{l}{\textit{Adjustment dynamics}} \\
$\tau_p$ & Price adjustment speed & 0.5  \\
$\tau_w$ & Technological adjustment speed & 0.5 \\
$\chi$ & Target inventory-to-output ratio & 5 \\
\midrule
\multicolumn{3}{l}{\textit{Production structure (nested CES, see Table~\ref{SM-tab:S_industries})}} \\
$\theta$ & Global curvature parameter & -2; (-3, -1)  \\
$\sigma$ & Global elasticity of substitution: $\sigma=1/(1-\theta)$ & 0.33; (0.25, 0.50) \\
$\sigma_{\mathrm{vadd}}$ & Value added: capital and labour nest & 1.25 \\
$\sigma_{\mathrm{energy}}$ & Energy sources nest & 1.67; (0.83, 2.50) \\
$\sigma_{\mathrm{digital}}$ & Digital industry nest & 1.25 \\
$\sigma_{\mathrm{other}}$ & Other nests: agriculture and forestry; services; manufacturing; fossil inputs; transport; waste; and mining & 0.50 \\
\multicolumn{3}{l}{\textit{Learning-by-doing in renewable technologies}} \\
$\mathrm{LR}_{\mathrm{hydro}}$
    & Learning rate for hydropower
    & $1.4\%$ \\
$\mathrm{LR}_{\mathrm{wind}}$
    & Learning rate for wind power
    & $7.0\%$ \\
$\mathrm{LR}_{\mathrm{biomass}}$
    & Learning rate for biomass
    & $5.0\%$ \\
$\mathrm{LR}_{\mathrm{solar}}$
    & Learning rate for solar photovoltaic power
    & $14.0\%$ \\
$\lambda_r$
    & Learning exponent implied by
      $\lambda_r=-\ln(1-\mathrm{LR}_r)/\ln(2)$
    & Hydro: $0.0203$; wind: $0.1047$;
      biomass: $0.0740$; solar photovoltaic: $0.2176$ \\
\midrule
\multicolumn{3}{l}{\textit{Households structure (nested CES, see Table~\ref{SM-tab:S_industries})}} \\   
$\theta'$ & Global curvature parameter & -2; (-3, -1)  \\
$\sigma'$ & Global elasticity of substitution: $\sigma'=1/(1-\theta')$ & 0.33; (0.25, 0.50) \\
$\sigma_{\mathrm{energy}}'$ & Energy sources nest & 1.67; (0.83, 2.50) \\
$\sigma_{\mathrm{digital}}'$ & Digital industry nest & 1.25 \\
$\sigma_{\mathrm{other}}'$ & Other nests: agriculture and food; services; manufacturing; fossil inputs; transport; waste; and mining & 0.50 \\
\midrule
\multicolumn{3}{l}{\textit{Labour market}} \\
$\tau_h$ & Labour mobility speed by skill with $h \in \{\text{low, medium, high}\}$ & 0.005; [0,1] \\
$\bar{s}$ & Similarity threshold & 0.25; (0.1, 0.75) \\
\midrule
\multicolumn{3}{l}{\textit{Green policy scenarios}} \\
Scenario & Climate policy regime (see Table~\ref{SM-tab:S_policy_scenarios}) & BAU, NDC, HWTP \\
\bottomrule
\end{tabular}
}
\end{table}
     
To assess the dependence of the aggregate results on the assumed substitution possibilities, Figures~\ref{SM-fig:global_elasticity_sensitivity} and~\ref{SM-fig:energy_elasticity_sensitivity} report sensitivity exercises for alternative values of the top-level elasticity, $\sigma$, and the elasticity within the energy nest, $\sigma_{\mathrm{energy}}$. The qualitative ordering of the policy scenarios is generally preserved, but the magnitude of the adjustment is parameter-sensitive, particularly under HWTP when substitution between energy inputs is severely constrained.

\section{Results}
\label{sec:results}

We now present the results of the decarbonisation scenarios. We focus on how much the simulated dynamics deviate from the BAU trajectory. Simulations are conducted over the period 2019--2035, with 2025 marking the end of the common calibration period shared by all scenarios. However, the model is run for an additional five years, keeping all simulation parameters fixed at their 2035 values, in order to verify that the observed dynamics are not driven by short-term transitional effects and that the main patterns remain stable beyond the main simulation horizon.

The results presented here correspond to a medium-run (benchmark in Table~\ref{tb:sim_parameters}) configuration in which partial factor mobility is allowed. At the upper level of the production function, the elasticity of substitution across nests is set to $\sigma = 0.33$, implying relatively limited substitution between broad input categories. Within nests, substitution possibilities are higher. The elasticity of substitution within the value-added nest, combining capital and labour, is set to $\sigma_{\mathrm{vadd}}=1.25$, while the energy nest, combining fossil and low-carbon energy inputs, is characterised by a higher elasticity of substitution, $\sigma_{\mathrm{energy}}=1.67$, allowing for a gradual reallocation between energy sources during the transition. All remaining intermediate-input nests are calibrated with a lower elasticity of substitution, $\sigma_{\mathrm{other}}=0.5$, with the exception of digital inputs, for which the elasticity of substitution is set to $\sigma_{\mathrm{digital}}=1.25$.\footnote{World-level aggregate results are presented in Figure~\ref{SM-fig:S_macro}, reporting the evolution of global carbon prices, emissions, output, GDP, prices, and household utility.}

\subsection{Sectoral gross value added dynamics}

\begin{figure}[h!]
\centering
\includegraphics[width=0.99\linewidth]{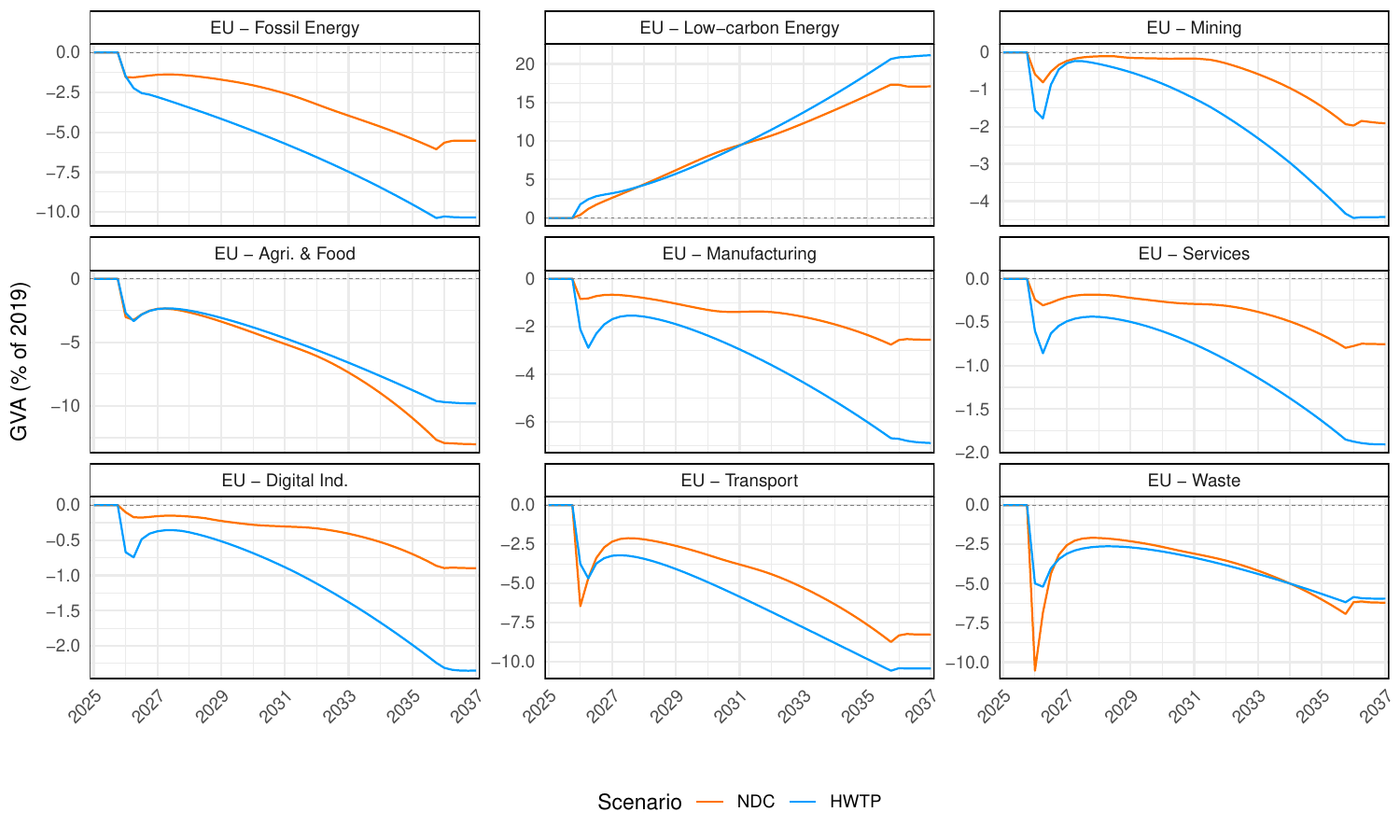}
\caption{Sectoral real GVA in the EU under the NDC and HWTP scenarios. Values are deviations from BAU expressed as a percentage of 2019 sectoral GVA.}
\label{fig:g_eu_gva}
\end{figure}

We start by discussing the sectoral reallocation in the EU. Figure~\ref{fig:g_eu_gva} reports the evolution of real gross value added (GVA) across broad economic sectors (see Table~\ref{SM-tab:S_industries}) as deviations from BAU, expressed as a percentage of 2019 values. From 2026, the most pronounced adjustments occur within the energy system.\footnote{Throughout the paper, the low-carbon energy category includes both mature and emerging renewable energy technologies, as well as other low-carbon technologies such as nuclear power.} Fossil energy activities experience a persistent contraction under both scenarios. By the end of the simulation period, GVA falls below BAU by amounts equivalent to approximately 5\% and 10\% of the sector's 2019 GVA under NDC and HWTP, respectively. By contrast, low-carbon energy GVA expands substantially, exceeding BAU by an amount equivalent to 17\% and 20\% of its 2019 level under NDC and HWTP, respectively.

Two factors contribute to this strong substitution between energy sources. First, the medium-run configuration allows sectors to partially adjust their input mix in response to changes in relative prices. Second, renewable energy sectors are calibrated to incorporate learning-by-doing effects, allowing productivity improvements to reinforce the expansion of alternative energy technologies over time \citep{rubin2015review}. Nevertheless, the rapid growth in GVA of low-carbon activities is far from implying a complete displacement of conventional energy sources. As we present below, fossil-based energy remains deeply embedded within production networks, limiting the speed at which low-carbon technologies can fully replace conventional sources.

Outside the energy system, changes in real GVA are also sizeable, reflecting the differences in sectors' exposure to climate transition according to the simulation scenarios. Agriculture and food, manufacturing, transport, and waste-management activities experience pronounced contractions. In the case of agriculture and waste, these effects are primarily associated with the emissions intensity of production processes rather than with fossil-fuel consumption itself. Manufacturing and transport can instead be comparatively more dependent on fossil energy inputs; therefore, they are particularly exposed to rising carbon prices. 

Services and digital industries display relatively small adjustments below BAU, less than 1\% under NDC and around 2\% under HWTP. Under the NDC scenario, these sectors are not directly subject to emissions taxation in the simulations, unlike HWTP, which is modelled as a sector-wide policy. Therefore, under NDC, the results reflect indirect equilibrium effects rather than direct policy exposure. 

Similarly, the mining sector experiences a relatively modest contraction in GVA, 2\% under NDC and more than 4\% under HWTP. Mining remains outside direct climate-policy coverage in both scenarios. Its contraction therefore also reflects indirect equilibrium effects, particularly the reduction in demand from fossil-energy activities.

Notably, sectoral impacts are larger in the HWTP scenario, except for the agriculture and food sector, where GVA falls slightly further under the NDC. The agriculture and food sector, which is covered by the EU ESR, has historically faced less stringent emissions targets than sectors in the EU ETS. This comparatively strong contraction is consistent with \cite{fellmann2018major}, which shows that stringent mitigation obligations in EU agriculture can induce substantial reductions in production when emissions abatement relies heavily on quantitative adjustments to activity levels.

\subsection{Production network recomposition}

We now examine how the transition changes the organisation of production networks. At each time $t$, the firm-to-firm component of the real input--output network is represented by the matrix $\mathbf Z_t=(z_{ij,t})_{i,j\in\mathcal{F}}$, where $z_{ij,t}=x_{ij,t}/y_{j,t}$ measures the amount of input supplied by firm $i$ per unit of real output produced by firm $j$. Because prices are normalised to one in 2019, real flows are expressed in units valued at 2019 prices. Using real coefficients prevents sectoral inflation from mechanically changing the network statistics and isolates changes in input requirements relative to real output. Unlike conventional input--output exercises in which technical coefficients are treated as fixed, these coefficients evolve endogenously as firms adjust their input mix in response to changes in relative prices. As such, they directly reflect firms' adjustments in input sourcing and the reallocation of intermediate input demand across suppliers induced by the transition.

One limitation of the model is that the binary structure of the production network is fixed: it does not account for the creation or destruction of supplier–customer relationships, but only for changes in the intensity of existing linkages. Consequently, the results capture functional recomposition rather than changes in network topology, since sectoral expansion and contraction can substantially modify sectors' economic roles and positions within production chains.

To investigate these mechanisms, we analyse a set of network statistics that capture changes in the importance of sectors as suppliers and users of intermediate inputs within production networks (see Section~\ref{SM-sec:S_network_measures} for network statistics definitions). Specifically, we focus on changes in in-strength and out-strength, Katz backward and forward centrality, and measures of upstreamness and downstreamness \citep{antras2012measuring}.

In-strength measures the total real intermediate-input requirement per unit of the buyer's real output, whereas out-strength measures how intensively a sector's output enters the direct production requirements of its customers. Backward and forward Katz centrality extend these measures to indirect dependencies: backward centrality captures reliance on upstream suppliers, while forward centrality captures a sector's reach through successive customer links. Finally, upstreamness measures distance from final demand and downstreamness measures distance from primary factors. Together, these statistics characterise how sectors are repositioned within production networks and how the functional organisation of production chains evolves during the transition.

\begin{figure}[h!]
\centering
\includegraphics[width=0.99\linewidth]{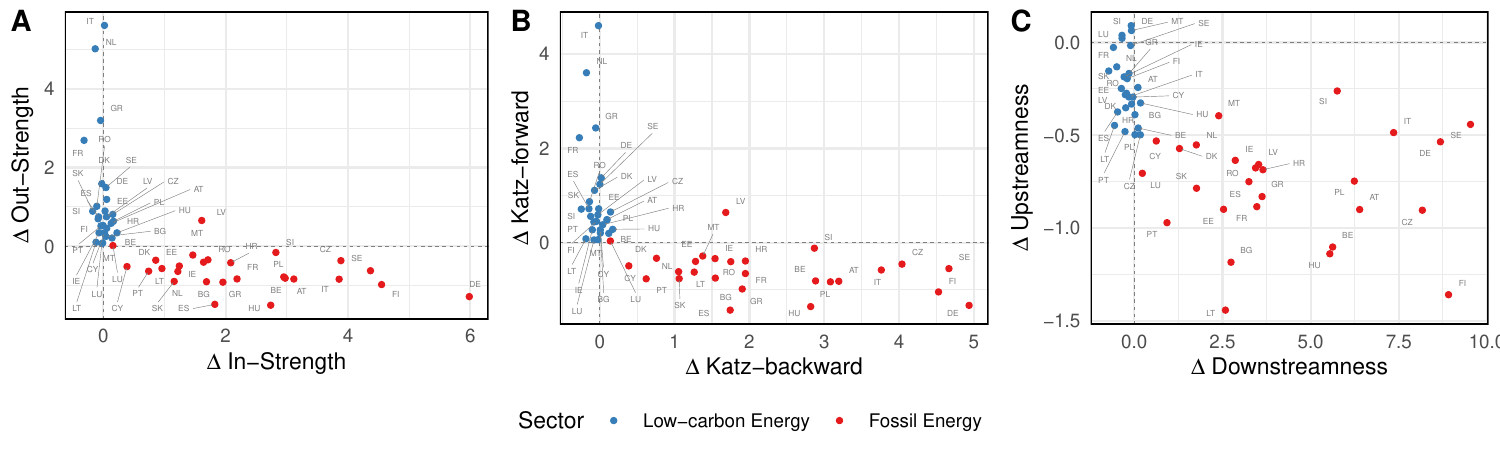}
\caption{Production-network adjustment in EU energy sectors under HWTP relative to BAU at the end of the simulation. Each point is a country-sector observation. All axes report HWTP--BAU differences in levels. Panel A plots in-strength against out-strength; Panel B backward against forward Katz centrality; and Panel C downstreamness against upstreamness.}
\label{fig:g_net_delta_eu}
\end{figure}

Figure~\ref{fig:g_net_delta_eu} reports country-level changes in production-network statistics for the energy sectors under HWTP relative to BAU. For each country-sector pair, we compute the HWTP--BAU difference at the end of the simulation for network centrality measures and indicators of sectoral position along production chains. Section~\ref{SM-sec:S_results} reports additional scatter plots of the corresponding statistics for all network nodes and effects at the sectoral level.

The results reveal an asymmetric repositioning of the two energy sectors within production networks. Fossil-energy sectors exhibit a clear increase in in-strength and backward Katz centrality, together with a marked increase in downstreamness. At the same time, out-strength declines in almost all EU countries, forward Katz centrality falls in most countries, and upstreamness decreases. Fossil activities therefore lose importance as direct suppliers and have less capacity to propagate shocks forward through buyer firms, while residual fossil production requires more intermediate inputs per unit of real output, depends more strongly on upstream suppliers, and occupies a relatively more downstream position.

Low-carbon energy sectors display a different pattern. Their out-strength and forward Katz centrality generally increase, indicating a stronger role as suppliers within the network, although the magnitude varies considerably across countries. Changes in in-strength and backward Katz centrality are smaller. Upstreamness declines moderately, while downstreamness changes little and without a uniform direction. These results indicate that the growing supplier role of low-carbon energy therefore does not reproduce symmetrically all the changes experienced by fossil energy.

Taken together, these patterns describe an asymmetric recomposition of the energy sectors within a fixed set of feasible production links. Fossil energy loses supplier relevance and forward influence but acquires stronger backward dependencies, while low-carbon energy becomes more important primarily as a supplier. The broader cross-sector results in Figure~\ref{SM-fig:S_net_sectors} show changes in linkage intensities and production-chain positions for particular sectors; the magnitudes of the differences from BAU are comparatively smaller than in the energy sector but still quite heterogeneous for some industries, exhibiting the network recompositions driven by indirect effects of the decarbonisation policy.

\subsection{Labour market adjustment}

Labour reallocation is a central mechanism through which the model captures the structural consequences of the green transition. Workers are allowed to move across sectors in response to wage differentials and industry-input similarity, but mobility is subject to adjustment frictions that reflect skill transferability and labour-market rigidities. 

\begin{figure}[h!]
\centering
\includegraphics[width=0.99\linewidth]{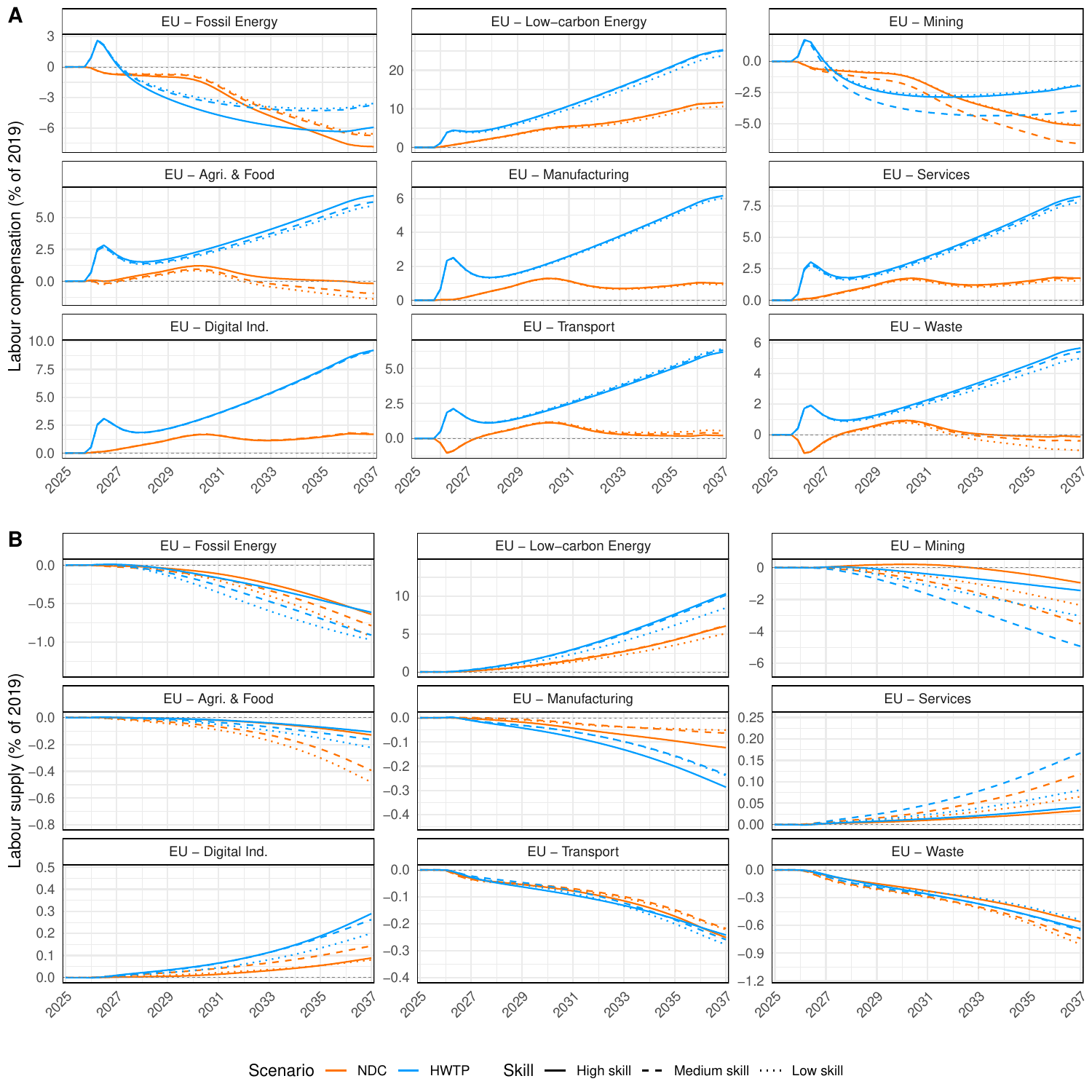}
\caption{Sectoral labour adjustment in the EU under the NDC and HWTP scenarios. Panel A reports total labour compensation and Panel B reports labour supply, both as deviations from BAU relative to their 2019 levels and by skill group.}
\label{fig:g_EU_labour}
\end{figure}

Figure~\ref{fig:g_EU_labour} reports total labour compensation and labour supply by economic sector and skill group as deviations from BAU, expressed as percentages of the corresponding 2019 levels. Here, the labour-adjustment parameter is set to $\tau_h=0.005$ for all workers, regardless of their skill level. This means that at most around 23\% of workers willing to move are able to do so in a simulation year (i.e., 52 time steps).

Changes in nominal labour compensation combine adjustments in hourly wages and hours employed, thereby capturing the effects of carbon pricing on sectoral labour expenditure. Figure~\ref{fig:g_EU_labour}A reveals heterogeneous adjustments across sectors. The most pronounced increase occurs in low-carbon energy, where labour compensation exceeds BAU by about 10\% under NDC and by more than 20\% under HWTP. Fossil-energy sectors experience persistent declines under both scenarios, with losses of approximately 3--8\%, depending on the scenario and skill group. Outside the energy system, HWTP produces broad increases in labour compensation across sectors. The adjustments are considerably smaller under NDC: compensation rises moderately in manufacturing, services, and digital industries, while differences from BAU remain close to zero or become slightly negative in agriculture and food, transport, and waste.

Across skill groups, labour compensation falls more sharply for high-skill workers in the fossil energy sector than for medium- and low-skill workers. Unlike fossil-energy industries, increases in labour compensation are relatively similar across skill groups, suggesting that the expansion of low-carbon activities generates broad-based labour demand rather than favouring a particular worker category. A connected negative spillover emerges in the mining sector, which is not directly regulated by climate policy yet reduces labour compensation as the use of fossil-energy inputs declines across the economy. 

Labour supply is measured as the available working hours allocated to each sector after labour reallocation. Figure~\ref{fig:g_EU_labour}B shows that the largest expansion occurs in low-carbon energy, reaching approximately 10\% of the corresponding 2019 allocation under HWTP and 6\% under NDC for some skill groups by the end of the displayed horizon. Mining experiences the strongest contraction, particularly among medium-skilled workers, with declines approaching 5\% under HWTP. Labour supply also decreases in fossil energy by approximately 0.6--1\% across scenarios and skill groups. Outside these sectors, deviations remain below 0.5\% in absolute terms, except in waste management, where reductions reach approximately 0.6--0.8\%. Services and digital industries register small increases in labour supply, while agriculture and food, manufacturing, and transport experience modest declines. All changes are measured relative to BAU and expressed as percentages of the corresponding 2019 labour allocations.

Overall, Figure~\ref{fig:g_EU_labour} reveals two distinct dimensions of labour-market adjustment. Changes in labour compensation are widespread across the economy and become particularly pronounced under HWTP. By contrast, the reallocation of available working hours is concentrated mainly in mining and energy-related activities and remains comparatively small in most other sectors. Outside these core sectors, the adjustment therefore operates more strongly through sectoral labour expenditure than through large movements in the available workforce.

\begin{figure}[h!]
\centering
\includegraphics[width=0.99\linewidth]{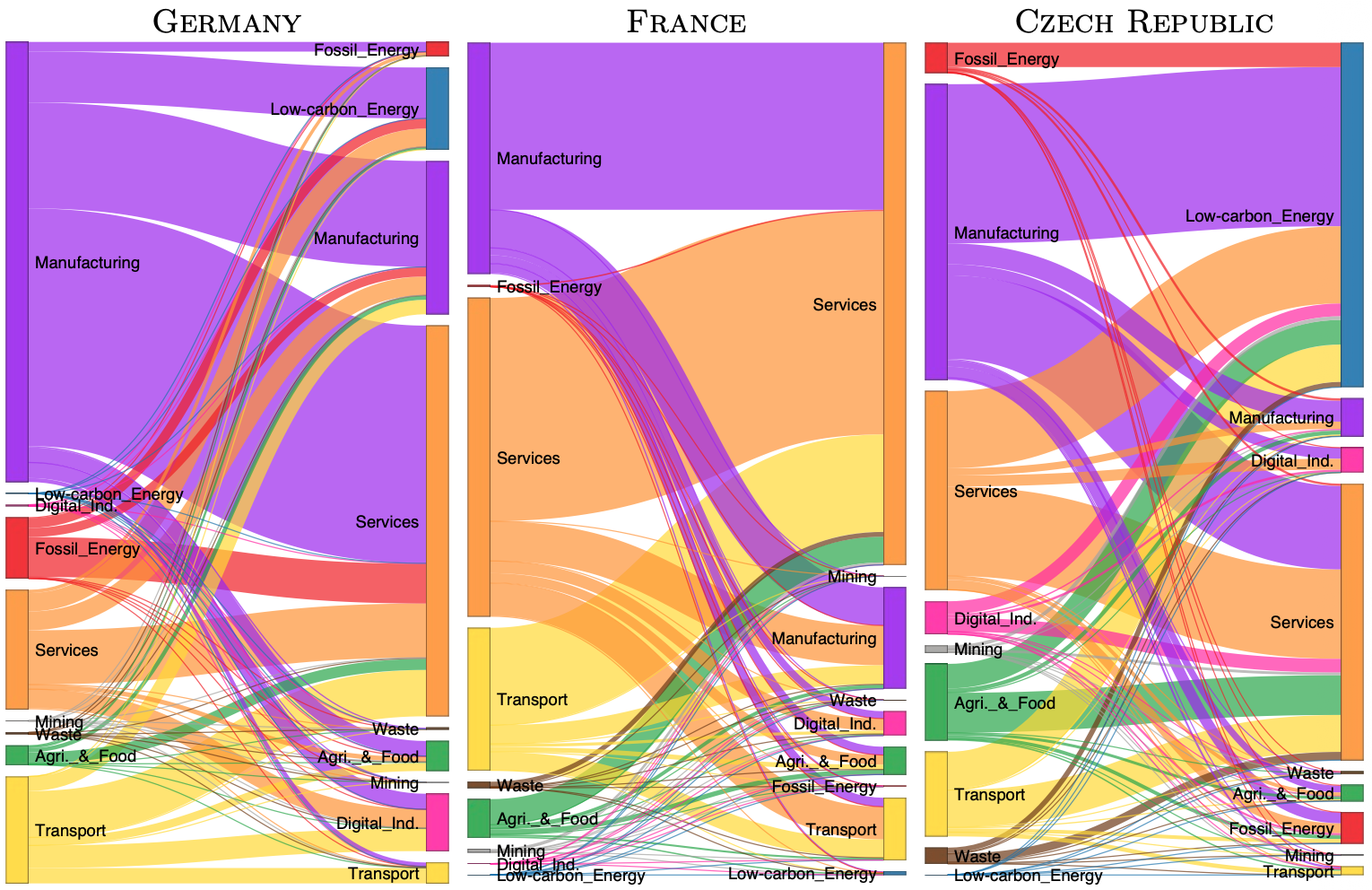}
\caption{Gross sectoral labour reallocation under the HWTP scenario. Sankey diagrams show flows of working hours across broad economic sectors from the initial year to the end of the simulation, with link widths proportional to the magnitude of labour reallocation. Results are shown for Germany, France, and the Czech Republic.}
\label{fig:g_sankey_labour}
\end{figure}

At the national level, the magnitude of labour reallocation depends on national production structures, energy mixes, and the scope for low-carbon energy expansion. To illustrate how this adjustment unfolds across countries, Figure~\ref{fig:g_sankey_labour} presents Sankey diagrams of the cumulative gross sectoral labour flows generated between 2019 and 2035 under the HWTP scenario for Germany, France, and the Czech Republic.\footnote{A Sankey plot represents the magnitude of flows between nodes---in this case, economic sectors---with the width of each link corresponding to the volume of workers moving from one sector to another.} Because labour markets in the model operate at the national level, the diagrams are constructed separately for each country and, unlike the other results in this section, these flows are not expressed as deviations from BAU.\footnote{They should therefore be interpreted as patterns of national labour reallocation occurring within the HWTP scenario, rather than as additional flows attributable exclusively to climate policy.}  

To start, the diagrams show that the transition does not necessarily entail a simple reallocation from fossil energy to low-carbon energy sectors, producing, instead, large flows between other sectors. We observe that labour flows tend to follow more complex adjustment paths, often involving movements towards broader sectors such as services or manufacturing. The services sector generally absorbs most labour reallocation across countries. Inflows from manufacturing, transport, and other sectors tend to be directed predominantly towards services.  

As seen in Figure~\ref{fig:g_EU_labour}B, the relative importance of labour outflows can vary markedly with the sector's initial size. For this reason, total inflows do not necessarily imply a large expansion of the services sector, since it is one of the largest employers in most European economies. However, sectors such as fossil energy or mining employ comparatively fewer workers in the baseline economy; hence, even moderate labour outflows can represent a significant adjustment relative to their initial employment. 

At the same time, the diagrams illustrate that sectoral adjustments can vary across countries. In the case of Germany and the Czech Republic, the enlargement of the available labour force in the low-carbon energy sector is comparatively large. In contrast, in the case of France, the aggregate flows do not suggest a pronounced contraction of fossil energy nor a substantial expansion of low-carbon energy sectors.\footnote{This does not imply that the transition has no effect in this country; rather, the representation reflects that net inflows and outflows do not scale sufficiently at the aggregate level relative to the impact in other sectors.}

The labour-market adjustments presented here show how firms shift expenditure to labour, while the available workforce gradually reallocates across sectors in response to wage differentials and mobility constraints. Because wages are sector-specific, some adjustment occurs through changes in remuneration rather than unemployment. Next, we complement the model outcomes with a reference-wage diagnostic that measures the labour reallocation required to close sectoral gaps.

\subsection{Labour gaps and mobility frictions}

In the model, the labour-market dynamics depend on the intrinsic frictions generated by the green transition, translated into changes in relative prices, the speed of the labour adjustment parameter and skill transferability. While the previous results assumed a benchmark value of the labour adjustment speed parameter, $\tau_h=0.005$, in practice, the labour relocation rate may be constrained by skill mismatches, relocation costs, information frictions, and barriers to occupational transitions. 

To consider the importance of these mechanisms, we conduct a sensitivity analysis that examines alternative assumptions about the speed of labour adjustment. Specifically, we add to the benchmark scenario cases characterised by maximum stickiness ($\tau_h = 0$) and rapid labour mobility ($\tau_h \rightarrow 1$). The objective is to quantify how labour-market frictions may amplify or mitigate potential unemployment during the transition.\footnote{Another source of variation comes from skill transferability. For the two complementary labour movement scenarios, we tested lower sparsity levels of the similarity matrix $s_{i,j}$ and obtained similar results, exposing the key role of wage differentials and the $\tau_h$ parameter.}

Consistent with the gradual adjustment mechanism described in the model, see Eq.\eqref{eq:labour_supply}, for each country-sector, skill group, and scenario, we measure the total labour gap relative to the corresponding country-skill average wage ($\Delta \ell^t_{h,i}$): positive values indicate sectoral excess labour supply (overcapacity), whereas negative values indicate sectoral labour shortages (vacancies). Because labour reallocation is defined within each country-skill group, aggregate overcapacities and vacancies offset each other. Their magnitude therefore indicates the potential transitional unemployment generated by the adjustment process in the presence of mobility frictions; in other words, the amount of labour that would need to be reallocated across sectors to restore balance. 

\begin{figure}[h!]
\centering
\includegraphics[width=0.99\linewidth]{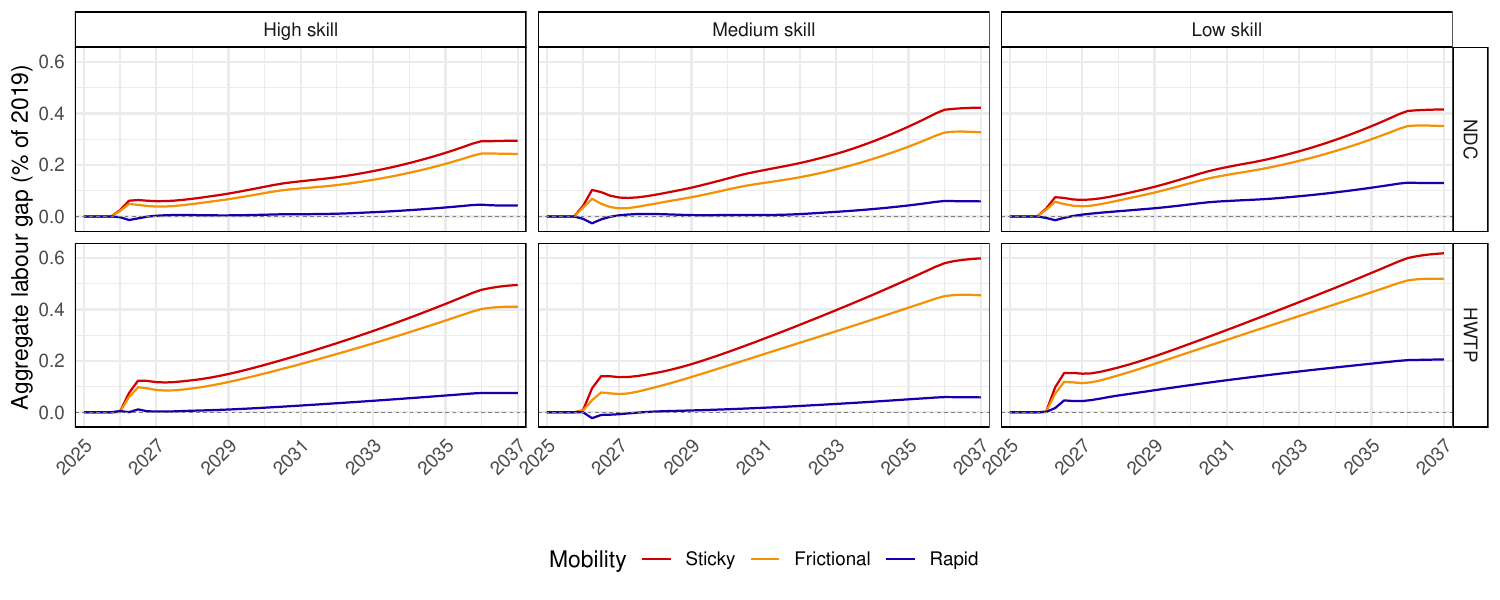}
\caption{Aggregate potential transitional unemployment in the EU under alternative speeds of labour mobility. Potential transitional unemployment is measured as the sum of positive sectoral reference-wage labour gaps, representing excess labour supply relative to the allocation implied by the corresponding country--skill average wage. Values report NDC and HWTP deviations from BAU relative to 2019 employment. Rows distinguish policy scenarios; columns distinguish skill groups; colours distinguish sticky ($\tau_h = 0$), frictional ($\tau_h = 0.005$), and rapid ($\tau_h = 0.95$) labour mobility.}
\label{fig:g_unemp_mob}
\end{figure}

Figure~\ref{fig:g_unemp_mob} reports the evolution of the potential transitional unemployment in the EU under alternative assumptions about labour mobility speed. Across all scenarios and skill groups, lower labour mobility is associated with substantially higher transitional unemployment. This is because when mobility is sticky, workers displaced from contracting sectors are unable to relocate rapidly towards expanding activities, generating persistent labour-market mismatches. By contrast, under rapid mobility, unemployment remains considerably lower throughout the simulation horizon, as workers can more easily respond to wage signals and move towards sectors experiencing labour shortages. 

Potential transitional unemployment can be systematically larger for low- and medium-skilled workers than for high-skilled workers, particularly under sticky and frictional mobility. Under the HWTP scenario, unemployment at the end of the simulation reaches about 0.6\% for low- and medium-skilled workers under sticky mobility, compared with around 0.5\% for high-skilled workers. The differences become particularly pronounced under rapid mobility: the additional gap for low-skilled workers remains close to 0.2\%, whereas it falls below 0.1\% for both medium- and high-skilled workers. A similar, although less pronounced, ordering emerges under the NDC scenario. Importantly, these differences do not arise from skill-specific mobility speed, which is held constant across skill groups and scenarios. They therefore reflect differences in the sectoral exposure and reallocation requirements of each skill class. Consequently, assuming that highly skilled workers move more easily across sectors, as might be expected in practice, could further widen this gap.

\begin{figure}[h!]
\centering 
\includegraphics[width=0.98\linewidth]{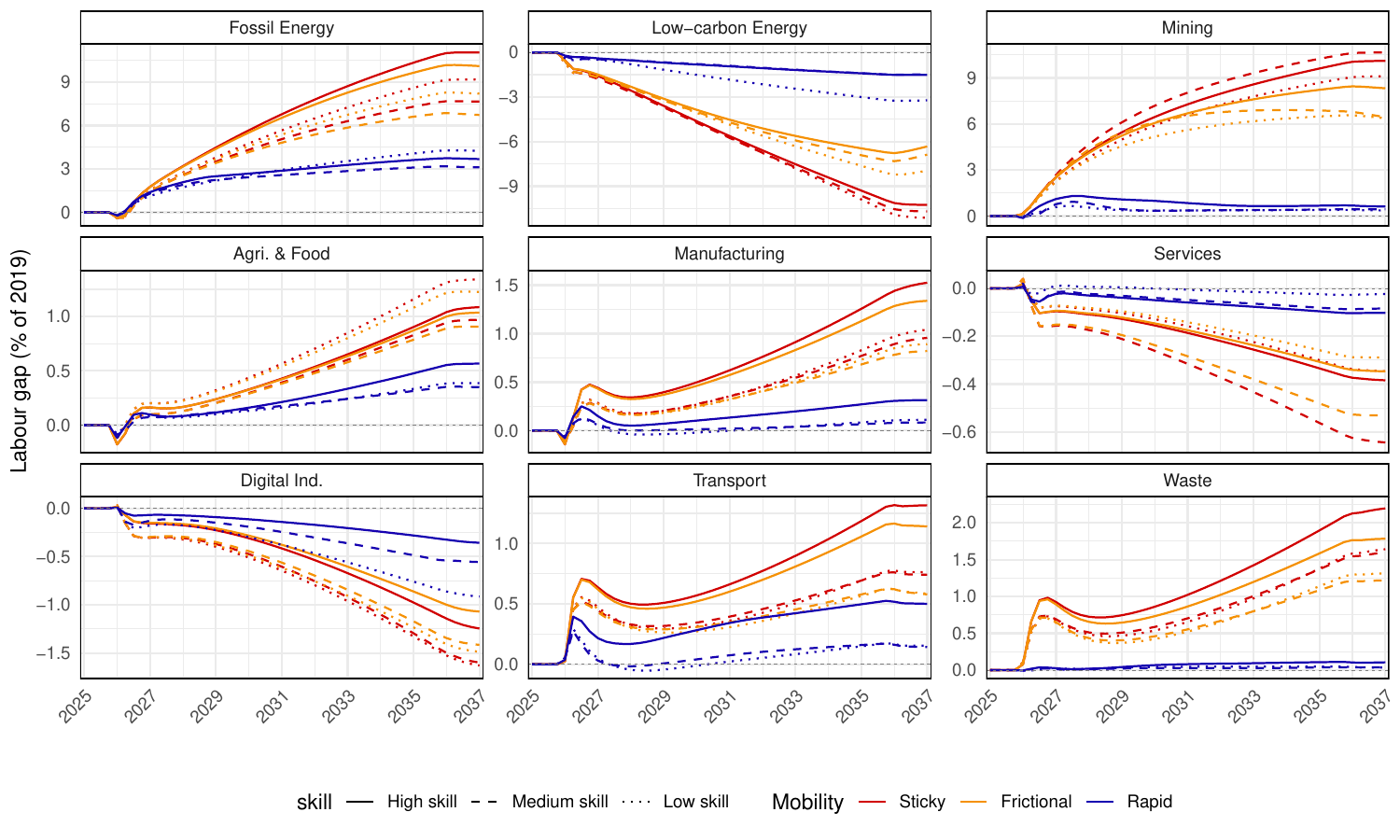}
\caption{Sectoral reference-wage labour gaps in the EU under alternative speeds of labour mobility. The gaps measure excess sectoral labour supply relative to the allocation implied by the corresponding country--skill average wage and are interpreted as potential transitional unemployment: positive values indicate labour overcapacity, and negative values indicate vacancies. Values report HWTP deviations from BAU relative to 2019 employment. Colours distinguish sticky ($\tau_h=0$), frictional ($\tau_h=0.005$), and rapid ($\tau_h=0.95$) labour mobility, while line types distinguish skill groups.}
\label{fig:g_unemp_mob_sec}
\end{figure}

To dig deeper into sectoral labour reallocation, we first aggregate labour gaps defined in Eq.~\eqref{eq:labour_gap} separately by sign at the broad sectoral level; consequently, the sum of positive values is total sectoral labour overcapacity, and the sum of negative values represents the number of sectoral vacancies. The values reported in Figure~\ref{fig:g_unemp_mob_sec} are HWTP deviations from the corresponding BAU values, expressed relative to each sector's 2019 employment. The largest share of potential transitional unemployment (positive values) emerges in fossil-energy and mining activities. Under sticky mobility, unemployment in fossil energy reaches around 10\% of its 2019 level by the end of the simulation horizon, while mining shows similar gaps for some skill groups. These effects are substantially attenuated when workers can move more rapidly across sectors. Under rapid mobility, the corresponding gaps remain only a few percentage points in fossil energy and close to zero in mining. The strong response of mining is particularly noteworthy because the sector is not directly targeted by the climate policy. Instead, its unemployment share reflects the contraction in production demand originating from fossil-energy activities, illustrating how the transition propagates upstream through production linkages.

The opposite pattern is observed in low-carbon energy activities. The expansion of low-carbon production generates transitional labour vacancies, which become particularly large when labour mobility is limited. Under sticky and frictional mobility, these vacancies increase progressively over the simulation horizon and reach shares comparable to the labour surpluses observed in fossil energy and mining. Rapid labour mobility substantially reduces these shortages.

The transition also produces smaller, but not negligible, labour gaps across the rest of the economy. Agriculture and food, manufacturing, transport, and waste management generally exhibit transitional unemployment, whereas digital industries and services display transitional vacancies. Although these effects are considerably smaller, in relative terms, than those observed in the energy and mining sectors, they show that the labour-market consequences of decarbonisation are not confined to the carbon-intensive activities directly affected. Rather, the adjustment involves a broader reallocation of workers across sectors connected through the production network.

Differences across skill groups vary by sector. Although the aggregate results indicated smaller labour gaps for high-skilled workers, the sectoral skill ordering is not uniform: in some activities high-skilled workers display relatively larger gaps, whereas in others the strongest gaps concern medium- or low-skilled workers. For instance, in sectors such as fossil energy, manufacturing, transport and waste, the higher overcapacity of high-skilled workers is more pronounced for scenarios with higher friction. Service sectors, by contrast, have more vacancies for medium-skilled workers. Since the simulations impose identical labour mobility speed across skill groups, these differences arise from their sectoral distribution and exposure to the transition rather than from an exogenously imposed mobility parametrisation favouring high-skilled workers.

Overall, the labour-market costs of the transition should not be interpreted exclusively as potential unemployment in declining sectors, since labour surpluses coexist with vacancies in expanding activities across the wider economy. Policies that facilitate worker mobility, retraining, and occupational transitions can therefore play a dual role: limiting persistent overcapacity in contracting carbon-intensive activities while helping expanding sectors fill transition-induced vacancies.

\section{Concluding remarks}  
\label{sec:concluding_remarks}

This paper examines the structural adjustment of production networks and sectoral labour market reallocation induced by climate policy. We develop a globally calibrated agent-based monetary input--output model in which climate policies change relative prices, intermediate-input demand, sectoral activity, and labour demand, while firms and workers adjust gradually. Focusing on the EU, we compare alternative decarbonisation scenarios with a BAU benchmark and trace how climate-policy shocks propagate through production networks and how skill-specific labour markets adjust. 

The simulations reveal substantial sectoral reallocation, with aggregate effects becoming markedly larger with policy stringency. The model is calibrated to the 2019 equilibrium and assumes fixed aggregate labour and capital endowments and no exogenous economy-wide trend-growth path, while allowing renewable-energy productivity to improve through learning-by-doing. Relative to BAU, global GHG emissions are approximately 27\% lower under NDC and 46\% lower under HWTP, whereas global GDP is approximately 2\% and 10\% lower, respectively. The implied GDP reductions are economically significant, particularly under HWTP, but remain smaller in percentage terms than the corresponding emissions reductions. Because BAU contains no exogenous trend-growth path, the GDP estimates represent differences in output levels relative to the same-year counterfactual rather than reductions in annual growth rates.\footnote{The qualitative ordering of the scenarios is preserved across the alternative values of the top-level elasticity considered in the sensitivity analysis. However, the magnitude of the adjustment depends on substitution possibilities within the energy system: under HWTP, very limited substitutability between energy inputs substantially amplifies the contraction in economic activity as shown in the SM.}

The strongest sectoral adjustments occur within the energy system: fossil-energy activities contract persistently, while low-carbon energy production expands. These effects propagate beyond directly regulated sectors through input--output linkages, affecting activities such as mining, manufacturing, transport, and agriculture.

The adjustment is particularly pronounced in sectoral GVA. In the EU, fossil-energy GVA falls below BAU by amounts equivalent to about 5\% and 10\% of its 2019 level under NDC and HWTP, respectively, while low-carbon energy GVA exceeds BAU by about 17\% and 20\% of its 2019 level. The effects also extend to suppliers of contracting activities, such as mining, even though the sector is not directly covered by carbon pricing in that scenario, reflecting the transmission of lower intermediate-input demand through production linkages.

The green transition produces an asymmetric recomposition of existing production-network linkages. Fossil-energy activities lose relevance as direct suppliers and exert less forward influence, while residual fossil production becomes more intensive in intermediate inputs per unit of output, more dependent on upstream suppliers, and relatively more downstream. Low-carbon energy sectors primarily strengthen their role as suppliers, with smaller and more heterogeneous changes in their backward dependencies and production-chain position. Because the binary set of input--output links remains fixed, these results describe changes in linkage intensities and sectoral positions within the existing production structure.

As regards the labour market adjustments induced by the transition, the simulations showed that adjustments occur through a broader set of intersectoral flows, with large sectors such as services absorbing a significant share of displaced labour, rather than an exclusive relocation from fossil to the low-carbon energy sector. 

While the policy stringency determines the extent of labour mismatches, the adjustment depends largely on labour-market frictions, which can trigger high sectoral transitional unemployment. When mobility is limited, relative contracting sectors can exhibit persistent labour overcapacity while relative expanding sectors face unfilled vacancies. Fossil energy and mining account for the largest labour surpluses, while low-carbon energy displays substantial labour shortages. Faster labour mobility compresses both sides of this imbalance, allowing workers released by declining activities to respond more rapidly to labour demand elsewhere in the economy. Importantly, differences across skill groups emerge even though mobility speed parameters are kept identical across skills, indicating that sectoral exposure and the structure of reallocation opportunities matter independently of assumed differences in workers' mobility.

From a policy perspective, this suggests that the costs of climate-policy adjustment depend not only on the stringency of environmental constraints but also on the economy's capacity to reallocate workers between declining and expanding activities. Policies that reduce barriers to intersectoral mobility may limit the persistence of transitional labour gaps. Although retraining is not modelled explicitly, it is one potential mechanism for reducing such barriers without altering the environmental objective itself.

\section*{Acknowledgements}
The authors gratefully acknowledge funding from the European Union's Horizon Europe research and innovation programme under grant agreement No.~101132559 (ST4TE--Strategies for Just and Equitable Transitions in Europe). The funding source had no role in the design of the study, the analysis or interpretation of the results, the preparation of the manuscript, or the decision to submit it for publication.

\clearpage
\begin{center}
\section*{\small Supplementary Material }
\vspace{10pt}
{\LARGE Labour reallocation and production-network recomposition in the green transition}
\end{center}
\vspace{10pt}

\setcounter{section}{0}
\setcounter{table}{0}
\setcounter{figure}{0}

\renewcommand{\thesection}{SM.\arabic{section}}
\renewcommand{\thetable}{SM.\arabic{table}}
\renewcommand{\thefigure}{SM.\arabic{figure}}
\renewcommand{\theHsection}{SM.\arabic{section}}
\renewcommand{\theHtable}{SM.\arabic{table}}
\renewcommand{\theHfigure}{SM.\arabic{figure}}


\section{Data and industry classifications}
\label{sec:S_data}

\subsection*{Production, consumption, and aggregate-sector classifications}

Table~\ref{tab:S_industries} presents the EXIOBASE industries, codes and names, together with the nesting structure used for both producers and households. Although the two classifications are highly similar, they are defined independently to reflect their distinct economic roles. For producers, nests represent groups of inputs within the production input mix, whereas for households, they represent categories within the consumption basket. The table also reports the economic-sector classification used throughout the analysis to aggregate industries into broader sectoral groups and facilitate the presentation and interpretation of the results.

\renewcommand{\arraystretch}{1.1}
\vspace{10pt}
{\tiny
\begin{longtable}{P{0.5cm}P{7cm}P{2.1cm}P{2.1cm}P{2.1cm}}
\caption{EXIOBASE industry classification, production and consumption nests, and aggregate economic sectors.}
\label{tab:S_industries}\label{SM-tab:S_industries}\\
\toprule
Code & Industry name & Firm nests & Household nests & Economic sector \\
\midrule
\endfirsthead
\toprule
Code & Industry name & Firm nests & Household nests & Economic sector \\
\midrule
\endhead
\bottomrule
\endfoot
1001 & Cultivation of paddy rice & Agri. \& Forestry & Agri. \& Food & Agri. \& Food \\
1002 & Cultivation of wheat & Agri. \& Forestry & Agri. \& Food & Agri. \& Food \\
1003 & Cultivation of cereal grains nec & Agri. \& Forestry & Agri. \& Food & Agri. \& Food \\
1004 & Cultivation of vegetables, fruit, nuts & Agri. \& Forestry & Agri. \& Food & Agri. \& Food \\
1005 & Cultivation of oil seeds & Agri. \& Forestry & Agri. \& Food & Agri. \& Food \\
1006 & Cultivation of sugar cane, sugar beet & Agri. \& Forestry & Agri. \& Food & Agri. \& Food \\
1007 & Cultivation of plant-based fibers & Agri. \& Forestry & Agri. \& Food & Agri. \& Food \\
1008 & Cultivation of crops nec & Agri. \& Forestry & Agri. \& Food & Agri. \& Food \\
1009 & Cattle farming & Agri. \& Forestry & Agri. \& Food & Agri. \& Food \\
1010 & Pigs farming & Agri. \& Forestry & Agri. \& Food & Agri. \& Food \\
1011 & Poultry farming & Agri. \& Forestry & Agri. \& Food & Agri. \& Food \\
1012 & Meat animals nec & Agri. \& Forestry & Agri. \& Food & Agri. \& Food \\
1013 & Animal products nec & Agri. \& Forestry & Agri. \& Food & Agri. \& Food \\
1014 & Raw milk & Agri. \& Forestry & Agri. \& Food & Agri. \& Food \\
1015 & Wool, silk-worm cocoons & Agri. \& Forestry & Agri. \& Food & Agri. \& Food \\
1016 & Manure treatment (conventional), storage and land application & Agri. \& Forestry & Agri. \& Food & Agri. \& Food \\
1017 & Manure treatment (biogas), storage and land application & Agri. \& Forestry & Agri. \& Food & Agri. \& Food \\
1018 & Forestry, logging and related service activities (02) & Agri. \& Forestry & Agri. \& Food & Agri. \& Food \\
1019 & Fishing, operating of fish hatcheries and fish farms; service activities incidental to fishing (05) & Agri. \& Forestry & Agri. \& Food & Agri. \& Food \\
1020 & Mining of coal and lignite; extraction of peat (10) & Fossil Input & Mining & Mining \\
1021 & Extraction of crude petroleum and services related to crude oil extraction, excluding surveying & Fossil Input & Mining & Mining \\
1022 & Extraction of natural gas and services related to natural gas extraction, excluding surveying & Fossil Input & Mining & Mining \\
1023 & Extraction, liquefaction, and regasification of other petroleum and gaseous materials & Fossil Input & Mining & Mining \\
1024 & Mining of uranium and thorium ores (12) & Mining & Mining & Mining \\
1025 & Mining of iron ores & Mining & Mining & Mining \\
1026 & Mining of copper ores and concentrates & Mining & Mining & Mining \\
1027 & Mining of nickel ores and concentrates & Mining & Mining & Mining \\
1028 & Mining of aluminium ores and concentrates & Mining & Mining & Mining \\
1029 & Mining of precious metal ores and concentrates & Mining & Mining & Mining \\
1030 & Mining of lead, zinc and tin ores and concentrates & Mining & Mining & Mining \\
1031 & Mining of other non-ferrous metal ores and concentrates & Mining & Mining & Mining \\
1032 & Quarrying of stone & Mining & Mining & Mining \\
1033 & Quarrying of sand and clay & Mining & Mining & Mining \\
1034 & Mining of chemical and fertilizer minerals, production of salt, other mining and quarrying n.e.c. & Mining & Mining & Mining \\
1035 & Processing of meat cattle & Agri. \& Forestry & Agri. \& Food & Agri. \& Food \\
1036 & Processing of meat pigs & Agri. \& Forestry & Agri. \& Food & Agri. \& Food \\
1037 & Processing of meat poultry & Agri. \& Forestry & Agri. \& Food & Agri. \& Food \\
1038 & Production of meat products nec & Agri. \& Forestry & Agri. \& Food & Agri. \& Food \\
1039 & Processing vegetable oils and fats & Manufacturing & Agri. \& Food & Agri. \& Food \\
1040 & Processing of dairy products & Agri. \& Forestry & Agri. \& Food & Agri. \& Food \\
1041 & Processed rice & Agri. \& Forestry & Agri. \& Food & Agri. \& Food \\
1042 & Sugar refining & Manufacturing & Agri. \& Food & Agri. \& Food \\
1043 & Processing of Food products nec & Agri. \& Forestry & Agri. \& Food & Agri. \& Food \\
1044 & Manufacture of beverages & Agri. \& Forestry & Agri. \& Food & Agri. \& Food \\
1045 & Manufacture of fish products & Agri. \& Forestry & Agri. \& Food & Agri. \& Food \\
1046 & Manufacture of tobacco products (16) & Agri. \& Forestry & Agri. \& Food & Agri. \& Food \\
1047 & Manufacture of textiles (17) & Manufacturing & Manufacturing & Manufacturing \\
1048 & Manufacture of wearing apparel; dressing and dyeing of fur (18) & Manufacturing & Manufacturing & Manufacturing \\
1049 & Tanning and dressing of leather; manufacture of luggage, handbags, saddlery, harness and footwear (19) & Manufacturing & Manufacturing & Manufacturing \\
1050 & Manufacture of wood and of products of wood and cork, except furniture; manufacture of articles of straw and plaiting materials (20) & Manufacturing & Manufacturing & Manufacturing \\
1051 & Re-processing of secondary wood material into new wood material & Manufacturing & Manufacturing & Manufacturing \\
1052 & Pulp & Manufacturing & Manufacturing & Manufacturing \\
1053 & Re-processing of secondary paper into new pulp & Manufacturing & Manufacturing & Manufacturing \\
1054 & Paper & Manufacturing & Manufacturing & Manufacturing \\
1055 & Publishing, printing and reproduction of recorded media (22) & Manufacturing & Manufacturing & Manufacturing \\
1056 & Manufacture of coke oven products & Energy & Energy & Fossil Energy \\
1057 & Petroleum Refinery & Energy & Energy & Fossil Energy \\
1058 & Processing of nuclear fuel & Energy & Energy & Low-carbon Energy \\
1059 & Plastics, basic & Manufacturing & Manufacturing & Manufacturing \\
1060 & Re-processing of secondary plastic into new plastic & Manufacturing & Manufacturing & Manufacturing \\
1061 & N-fertiliser & Manufacturing & Manufacturing & Manufacturing \\
1062 & P- and other fertiliser & Manufacturing & Manufacturing & Manufacturing \\
1063 & Chemicals nec & Manufacturing & Manufacturing & Manufacturing \\
1064 & Manufacture of rubber and plastic products (25) & Manufacturing & Manufacturing & Manufacturing \\
1065 & Manufacture of glass and glass products & Manufacturing & Manufacturing & Manufacturing \\
1066 & Re-processing of secondary glass into new glass & Manufacturing & Manufacturing & Manufacturing \\
1067 & Manufacture of ceramic goods & Manufacturing & Manufacturing & Manufacturing \\
1068 & Manufacture of bricks, tiles and construction products, in baked clay & Manufacturing & Manufacturing & Manufacturing \\
1069 & Manufacture of cement, lime and plaster & Manufacturing & Manufacturing & Manufacturing \\
1070 & Re-processing of ash into clinker & Manufacturing & Manufacturing & Manufacturing \\
1071 & Manufacture of other non-metallic mineral products n.e.c. & Manufacturing & Manufacturing & Manufacturing \\
1072 & Manufacture of basic iron and steel and of ferro-alloys and first products thereof & Manufacturing & Manufacturing & Manufacturing \\
1073 & Re-processing of secondary steel into new steel & Manufacturing & Manufacturing & Manufacturing \\
1074 & Precious metals production & Manufacturing & Manufacturing & Manufacturing \\
1075 & Re-processing of secondary preciuos metals into new preciuos metals & Manufacturing & Manufacturing & Manufacturing \\
1076 & Aluminium production & Manufacturing & Manufacturing & Manufacturing \\
1077 & Re-processing of secondary aluminium into new aluminium & Manufacturing & Manufacturing & Manufacturing \\
1078 & Lead, zinc and tin production & Manufacturing & Manufacturing & Manufacturing \\
1079 & Re-processing of secondary lead into new lead, zinc and tin & Manufacturing & Manufacturing & Manufacturing \\
1080 & Copper production & Manufacturing & Manufacturing & Manufacturing \\
1081 & Re-processing of secondary copper into new copper & Manufacturing & Manufacturing & Manufacturing \\
1082 & Other non-ferrous metal production & Manufacturing & Manufacturing & Manufacturing \\
1083 & Re-processing of secondary other non-ferrous metals into new other non-ferrous metals & Manufacturing & Manufacturing & Manufacturing \\
1084 & Casting of metals & Manufacturing & Manufacturing & Manufacturing \\
1085 & Manufacture of fabricated metal products, except machinery and equipment (28) & Manufacturing & Manufacturing & Manufacturing \\
1086 & Manufacture of machinery and equipment n.e.c. (29) & Manufacturing & Manufacturing & Manufacturing \\
1087 & Manufacture of office machinery and computers (30) & Digital Ind. & Digital Ind. & Digital Ind. \\
1088 & Manufacture of electrical machinery and apparatus n.e.c. (31) & Digital Ind. & Digital Ind. & Digital Ind. \\
1089 & Manufacture of radio, television and communication equipment and apparatus (32) & Digital Ind. & Digital Ind. & Digital Ind. \\
1090 & Manufacture of medical, precision and optical instruments, watches and clocks (33) & Digital Ind. & Digital Ind. & Digital Ind. \\
1091 & Manufacture of motor vehicles, trailers and semi-trailers (34) & Manufacturing & Manufacturing & Manufacturing \\
1092 & Manufacture of other transport equipment (35) & Manufacturing & Manufacturing & Manufacturing \\
1093 & Manufacture of furniture; manufacturing n.e.c. (36) & Manufacturing & Manufacturing & Manufacturing \\
1094 & Recycling of waste and scrap & Manufacturing & Manufacturing & Manufacturing \\
1095 & Recycling of bottles by direct reuse & Manufacturing & Manufacturing & Manufacturing \\
1096 & Production of electricity by coal & Energy & Energy & Fossil Energy \\
1097 & Production of electricity by gas & Energy & Energy & Fossil Energy \\
1098 & Production of electricity by nuclear & Energy & Energy & Low-carbon Energy \\
1099 & Production of electricity by hydro & Energy & Energy & Low-carbon Energy \\
1100 & Production of electricity by wind & Energy & Energy & Low-carbon Energy \\
1101 & Production of electricity by petroleum and other oil derivatives & Energy & Energy & Fossil Energy \\
1102 & Production of electricity by biomass and waste & Energy & Energy & Low-carbon Energy \\
1103 & Production of electricity by solar photovoltaic & Energy & Energy & Low-carbon Energy \\
1104 & Production of electricity by solar thermal & Energy & Energy & Low-carbon Energy \\
1105 & Production of electricity by tide, wave, ocean & Energy & Energy & Low-carbon Energy \\
1106 & Production of electricity by Geothermal & Energy & Energy & Low-carbon Energy \\
1107 & Production of electricity nec & Energy & Energy & Fossil Energy \\
1108 & Transmission of electricity & Manufacturing & Services & Services \\
1109 & Distribution and trade of electricity & Services & Services & Services \\
1110 & Manufacture of gas; distribution of gaseous fuels through mains & Energy & Energy & Fossil Energy \\
1111 & Steam and hot water supply & Energy & Energy & Fossil Energy \\
1112 & Collection, purification and distribution of water (41) & Services & Services & Services \\
1113 & Construction (45) & Services & Services & Services \\
1114 & Re-processing of secondary construction material into aggregates & Services & Services & Services \\
1115 & Sale, maintenance, repair of motor vehicles, motor vehicles parts, motorcycles, motor cycles parts and accessoiries & Services & Services & Services \\
1116 & Retail sale of automotive fuel & Energy & Energy & Fossil Energy \\
1117 & Wholesale trade and commission trade, except of motor vehicles and motorcycles (51) & Services & Services & Services \\
1118 & Retail trade, except of motor vehicles and motorcycles; repair of personal and household goods (52) & Services & Services & Services \\
1119 & Hotels and restaurants (55) & Services & Services & Services \\
1120 & Transport via railways & Services & Transport & Transport \\
1121 & Other land transport & Services & Transport & Transport \\
1122 & Transport via pipelines & Services & Manufacturing & Transport \\
1123 & Sea and coastal water transport & Services & Transport & Transport \\
1124 & Inland water transport & Services & Transport & Transport \\
1125 & Air transport (62) & Services & Transport & Transport \\
1126 & Supporting and auxiliary transport activities; activities of travel agencies (63) & Services & Services & Services \\
1127 & Post and telecommunications (64) & Digital Ind. & Digital Ind. & Digital Ind. \\
1128 & Financial intermediation, except insurance and pension funding (65) & Services & Services & Services \\
1129 & Insurance and pension funding, except compulsory social security (66) & Services & Services & Services \\
1130 & Activities auxiliary to financial intermediation (67) & Services & Services & Services \\
1131 & Real estate activities (70) & Services & Services & Services \\
1132 & Renting of machinery and equipment without operator and of personal and household goods (71) & Services & Services & Services \\
1133 & Computer and related activities (72) & Digital Ind. & Digital Ind. & Digital Ind. \\
1134 & Research and development (73) & Services & Services & Services \\
1135 & Other business activities (74) & Services & Services & Services \\
1136 & Public administration and defence; compulsory social security (75) & Services & Services & Services \\
1137 & Education (80) & Services & Services & Services \\
1138 & Health and social work (85) & Services & Services & Services \\
1139 & Incineration of waste: Food & Services & Manufacturing & Manufacturing \\
1140 & Incineration of waste: Paper & Services & Manufacturing & Manufacturing \\
1141 & Incineration of waste: Plastic & Services & Manufacturing & Manufacturing \\
1142 & Incineration of waste: Metals and Inert materials & Services & Manufacturing & Manufacturing \\
1143 & Incineration of waste: Textiles & Services & Manufacturing & Manufacturing \\
1144 & Incineration of waste: Wood & Services & Manufacturing & Manufacturing \\
1145 & Incineration of waste: Oil/Hazardous waste & Services & Manufacturing & Manufacturing \\
1146 & Biogasification of food waste, incl. land application & Services & Waste & Waste \\
1147 & Biogasification of paper, incl. land application & Services & Waste & Waste \\
1148 & Biogasification of sewage slugde, incl. land application & Services & Waste & Waste \\
1149 & Composting of food waste, incl. land application & Services & Waste & Waste \\
1150 & Composting of paper and wood, incl. land application & Services & Waste & Waste \\
1151 & Waste water treatment, food & Services & Waste & Waste \\
1152 & Waste water treatment, other & Services & Waste & Waste \\
1153 & Landfill of waste: Food & Services & Waste & Waste \\
1154 & Landfill of waste: Paper & Services & Waste & Waste \\
1155 & Landfill of waste: Plastic & Services & Waste & Waste \\
1156 & Landfill of waste: Inert/metal/hazardous & Services & Waste & Waste \\
1157 & Landfill of waste: Textiles & Services & Waste & Waste \\
1158 & Landfill of waste: Wood & Services & Waste & Waste \\
1159 & Activities of membership organisation n.e.c. (91) & Services & Services & Services \\
1160 & Recreational, cultural and sporting activities (92) & Services & Services & Services \\
1161 & Other service activities (93) & Services & Services & Services \\
1162 & Private households with employed persons (95) & Services & Services & Services \\
1163 & Extra-territorial organizations and bodies & Services & Services & Services \\
\end{longtable}
}

\subsection*{Climate-policy sector coverage}

Table~\ref{tab:S_policy_industries} reports the classification of EXIOBASE industries used in the implementation of green policies. For each industry, the table provides indicators identifying whether the industry is classified as a fossil-input sector, covered by the European Union Emissions Trading System (EU ETS) up to 2025, or included within the sectors regulated under the European Union Effort Sharing Regulation (EU ESR). 
\renewcommand{\arraystretch}{1.1}
\vspace{10pt}
{\tiny
\begin{longtable}{P{0.5cm}P{7cm}P{1.3cm}P{1cm}P{1cm}P{2.1cm}}
\caption{Industry coverage of fossil inputs, the EU Emissions Trading System, and the EU Effort Sharing Regulation.}
\label{tab:S_policy_industries}\label{SM-tab:S_policy_industries}\\
\toprule
Code & Industry name & Fossil input & EU-ETS &  EU-ESR & Policy sector \\
\midrule
\endfirsthead
\toprule
Code & Industry name & Fossil input & EU-ETS &  EU-ESR & Policy sector \\
\midrule
\endhead
\bottomrule
\endfoot
1001 & Cultivation of paddy rice & 0 & 0 & 1 & Agri. \& Forestry \\
1002 & Cultivation of wheat & 0 & 0 & 1 & Agri. \& Forestry \\
1003 & Cultivation of cereal grains nec & 0 & 0 & 1 & Agri. \& Forestry \\
1004 & Cultivation of vegetables, fruit, nuts & 0 & 0 & 1 & Agri. \& Forestry \\
1005 & Cultivation of oil seeds & 0 & 0 & 1 & Agri. \& Forestry \\
1006 & Cultivation of sugar cane, sugar beet & 0 & 0 & 1 & Agri. \& Forestry \\
1007 & Cultivation of plant-based fibers & 0 & 0 & 1 & Agri. \& Forestry \\
1008 & Cultivation of crops nec & 0 & 0 & 1 & Agri. \& Forestry \\
1009 & Cattle farming & 0 & 0 & 1 & Agri. \& Forestry \\
1010 & Pigs farming & 0 & 0 & 1 & Agri. \& Forestry \\
1011 & Poultry farming & 0 & 0 & 1 & Agri. \& Forestry \\
1012 & Meat animals nec & 0 & 0 & 1 & Agri. \& Forestry \\
1013 & Animal products nec & 0 & 0 & 1 & Agri. \& Forestry \\
1014 & Raw milk & 0 & 0 & 1 & Agri. \& Forestry \\
1015 & Wool, silk-worm cocoons & 0 & 0 & 1 & Agri. \& Forestry \\
1016 & Manure treatment (conventional), storage and land application & 0 & 0 & 1 & Agri. \& Forestry \\
1017 & Manure treatment (biogas), storage and land application & 0 & 0 & 1 & Agri. \& Forestry \\
1018 & Forestry, logging and related service activities (02) & 0 & 0 & 1 & Agri. \& Forestry \\
1019 & Fishing, operating of fish hatcheries and fish farms; service activities incidental to fishing (05) & 0 & 0 & 1 & Agri. \& Forestry \\
1020 & Mining of coal and lignite; extraction of peat (10) & 1 & 0 & 0 & Fossil Input \\
1021 & Extraction of crude petroleum and services related to crude oil extraction, excluding surveying & 1 & 0 & 0 & Fossil Input \\
1022 & Extraction of natural gas and services related to natural gas extraction, excluding surveying & 1 & 0 & 0 & Fossil Input \\
1023 & Extraction, liquefaction, and regasification of other petroleum and gaseous materials & 1 & 0 & 0 & Fossil Input \\
1024 & Mining of uranium and thorium ores (12) & 0 & 0 & 0 & Mining \\
1025 & Mining of iron ores & 0 & 0 & 0 & Mining \\
1026 & Mining of copper ores and concentrates & 0 & 0 & 0 & Mining \\
1027 & Mining of nickel ores and concentrates & 0 & 0 & 0 & Mining \\
1028 & Mining of aluminium ores and concentrates & 0 & 0 & 0 & Mining \\
1029 & Mining of precious metal ores and concentrates & 0 & 0 & 0 & Mining \\
1030 & Mining of lead, zinc and tin ores and concentrates & 0 & 0 & 0 & Mining \\
1031 & Mining of other non-ferrous metal ores and concentrates & 0 & 0 & 0 & Mining \\
1032 & Quarrying of stone & 0 & 0 & 0 & Mining \\
1033 & Quarrying of sand and clay & 0 & 0 & 0 & Mining \\
1034 & Mining of chemical and fertilizer minerals, production of salt, other mining and quarrying n.e.c. & 0 & 0 & 0 & Mining \\
1035 & Processing of meat cattle & 0 & 0 & 0 & Other \\
1036 & Processing of meat pigs & 0 & 0 & 0 & Other \\
1037 & Processing of meat poultry & 0 & 0 & 0 & Other \\
1038 & Production of meat products nec & 0 & 0 & 0 & Other \\
1039 & Processing vegetable oils and fats & 0 & 1 & 0 & Manufacturing \\
1040 & Processing of dairy products & 0 & 0 & 0 & Other \\
1041 & Processed rice & 0 & 0 & 0 & Other \\
1042 & Sugar refining & 0 & 1 & 0 & Manufacturing \\
1043 & Processing of Food products nec & 0 & 0 & 0 & Other \\
1044 & Manufacture of beverages & 0 & 0 & 0 & Other \\
1045 & Manufacture of fish products & 0 & 0 & 0 & Other \\
1046 & Manufacture of tobacco products (16) & 0 & 0 & 0 & Other \\
1047 & Manufacture of textiles (17) & 0 & 1 & 0 & Manufacturing \\
1048 & Manufacture of wearing apparel; dressing and dyeing of fur (18) & 0 & 1 & 0 & Manufacturing \\
1049 & Tanning and dressing of leather; manufacture of luggage, handbags, saddlery, harness and footwear (19) & 0 & 1 & 0 & Manufacturing \\
1050 & Manufacture of wood and of products of wood and cork, except furniture; manufacture of articles of straw and plaiting materials (20) & 0 & 1 & 0 & Manufacturing \\
1051 & Re-processing of secondary wood material into new wood material & 0 & 1 & 0 & Manufacturing \\
1052 & Pulp & 0 & 1 & 0 & Manufacturing \\
1053 & Re-processing of secondary paper into new pulp & 0 & 1 & 0 & Manufacturing \\
1054 & Paper & 0 & 1 & 0 & Manufacturing \\
1055 & Publishing, printing and reproduction of recorded media (22) & 0 & 0 & 0 & Other \\
1056 & Manufacture of coke oven products & 1 & 1 & 0 & Electricity \& Heat \\
1057 & Petroleum Refinery & 1 & 1 & 0 & Electricity \& Heat \\
1058 & Processing of nuclear fuel & 0 & 0 & 0 & Other \\
1059 & Plastics, basic & 0 & 1 & 0 & Manufacturing \\
1060 & Re-processing of secondary plastic into new plastic & 0 & 1 & 0 & Manufacturing \\
1061 & N-fertiliser & 0 & 1 & 0 & Manufacturing \\
1062 & P- and other fertiliser & 0 & 1 & 0 & Manufacturing \\
1063 & Chemicals nec & 0 & 1 & 0 & Manufacturing \\
1064 & Manufacture of rubber and plastic products (25) & 0 & 1 & 0 & Manufacturing \\
1065 & Manufacture of glass and glass products & 0 & 1 & 0 & Manufacturing \\
1066 & Re-processing of secondary glass into new glass & 0 & 1 & 0 & Manufacturing \\
1067 & Manufacture of ceramic goods & 0 & 1 & 0 & Manufacturing \\
1068 & Manufacture of bricks, tiles and construction products, in baked clay & 0 & 1 & 0 & Manufacturing \\
1069 & Manufacture of cement, lime and plaster & 0 & 1 & 0 & Manufacturing \\
1070 & Re-processing of ash into clinker & 0 & 1 & 0 & Manufacturing \\
1071 & Manufacture of other non-metallic mineral products n.e.c. & 0 & 1 & 0 & Manufacturing \\
1072 & Manufacture of basic iron and steel and of ferro-alloys and first products thereof & 0 & 1 & 0 & Manufacturing \\
1073 & Re-processing of secondary steel into new steel & 0 & 1 & 0 & Manufacturing \\
1074 & Precious metals production & 0 & 0 & 0 & Other \\
1075 & Re-processing of secondary preciuos metals into new preciuos metals & 0 & 0 & 0 & Other \\
1076 & Aluminium production & 0 & 1 & 0 & Manufacturing \\
1077 & Re-processing of secondary aluminium into new aluminium & 0 & 1 & 0 & Manufacturing \\
1078 & Lead, zinc and tin production & 0 & 1 & 0 & Manufacturing \\
1079 & Re-processing of secondary lead into new lead, zinc and tin & 0 & 1 & 0 & Manufacturing \\
1080 & Copper production & 0 & 1 & 0 & Manufacturing \\
1081 & Re-processing of secondary copper into new copper & 0 & 1 & 0 & Manufacturing \\
1082 & Other non-ferrous metal production & 0 & 1 & 0 & Manufacturing \\
1083 & Re-processing of secondary other non-ferrous metals into new other non-ferrous metals & 0 & 1 & 0 & Manufacturing \\
1084 & Casting of metals & 0 & 1 & 0 & Manufacturing \\
1085 & Manufacture of fabricated metal products, except machinery and equipment (28) & 0 & 0 & 0 & Other \\
1086 & Manufacture of machinery and equipment n.e.c. (29) & 0 & 0 & 0 & Other \\
1087 & Manufacture of office machinery and computers (30) & 0 & 0 & 0 & Other \\
1088 & Manufacture of electrical machinery and apparatus n.e.c. (31) & 0 & 0 & 0 & Other \\
1089 & Manufacture of radio, television and communication equipment and apparatus (32) & 0 & 0 & 0 & Other \\
1090 & Manufacture of medical, precision and optical instruments, watches and clocks (33) & 0 & 0 & 0 & Other \\
1091 & Manufacture of motor vehicles, trailers and semi-trailers (34) & 0 & 0 & 0 & Other \\
1092 & Manufacture of other transport equipment (35) & 0 & 1 & 0 & Manufacturing \\
1093 & Manufacture of furniture; manufacturing n.e.c. (36) & 0 & 0 & 0 & Other \\
1094 & Recycling of waste and scrap & 0 & 0 & 0 & Other \\
1095 & Recycling of bottles by direct reuse & 0 & 0 & 0 & Other \\
1096 & Production of electricity by coal & 1 & 1 & 0 & Electricity \& Heat \\
1097 & Production of electricity by gas & 1 & 1 & 0 & Electricity \& Heat \\
1098 & Production of electricity by nuclear & 0 & 0 & 0 & Other \\
1099 & Production of electricity by hydro & 0 & 0 & 0 & Other \\
1100 & Production of electricity by wind & 0 & 0 & 0 & Other \\
1101 & Production of electricity by petroleum and other oil derivatives & 1 & 1 & 0 & Electricity \& Heat \\
1102 & Production of electricity by biomass and waste & 0 & 0 & 0 & Other \\
1103 & Production of electricity by solar photovoltaic & 0 & 0 & 0 & Other \\
1104 & Production of electricity by solar thermal & 0 & 0 & 0 & Other \\
1105 & Production of electricity by tide, wave, ocean & 0 & 0 & 0 & Other \\
1106 & Production of electricity by Geothermal & 0 & 0 & 0 & Other \\
1107 & Production of electricity nec & 0 & 1 & 0 & Electricity \& Heat \\
1108 & Transmission of electricity & 0 & 0 & 0 & Other \\
1109 & Distribution and trade of electricity & 0 & 0 & 0 & Other \\
1110 & Manufacture of gas; distribution of gaseous fuels through mains & 1 & 1 & 0 & Electricity \& Heat \\
1111 & Steam and hot water supply & 0 & 1 & 0 & Electricity \& Heat \\
1112 & Collection, purification and distribution of water (41) & 0 & 0 & 0 & Other \\
1113 & Construction (45) & 0 & 1 & 0 & Manufacturing \\
1114 & Re-processing of secondary construction material into aggregates & 0 & 0 & 0 & Other \\
1115 & Sale, maintenance, repair of motor vehicles, motor vehicles parts, motorcycles, motor cycles parts and accessoiries & 0 & 0 & 0 & Other \\
1116 & Retail sale of automotive fuel & 1 & 0 & 0 & Fossil Input \\
1117 & Wholesale trade and commission trade, except of motor vehicles and motorcycles (51) & 0 & 0 & 0 & Other \\
1118 & Retail trade, except of motor vehicles and motorcycles; repair of personal and household goods (52) & 0 & 0 & 0 & Other \\
1119 & Hotels and restaurants (55) & 0 & 0 & 0 & Other \\
1120 & Transport via railways & 0 & 0 & 1 & Transport \\
1121 & Other land transport & 0 & 0 & 1 & Transport \\
1122 & Transport via pipelines & 0 & 1 & 0 & Manufacturing \\
1123 & Sea and coastal water transport & 0 & 1 & 0 & Transport \\
1124 & Inland water transport & 0 & 0 & 1 & Transport \\
1125 & Air transport (62) & 0 & 1 & 0 & Transport \\
1126 & Supporting and auxiliary transport activities; activities of travel agencies (63) & 0 & 0 & 0 & Other \\
1127 & Post and telecommunications (64) & 0 & 0 & 0 & Other \\
1128 & Financial intermediation, except insurance and pension funding (65) & 0 & 0 & 0 & Other \\
1129 & Insurance and pension funding, except compulsory social security (66) & 0 & 0 & 0 & Other \\
1130 & Activities auxiliary to financial intermediation (67) & 0 & 0 & 0 & Other \\
1131 & Real estate activities (70) & 0 & 0 & 0 & Other \\
1132 & Renting of machinery and equipment without operator and of personal and household goods (71) & 0 & 0 & 0 & Other \\
1133 & Computer and related activities (72) & 0 & 0 & 0 & Other \\
1134 & Research and development (73) & 0 & 0 & 0 & Other \\
1135 & Other business activities (74) & 0 & 0 & 0 & Other \\
1136 & Public administration and defence; compulsory social security (75) & 0 & 0 & 0 & Other \\
1137 & Education (80) & 0 & 0 & 0 & Other \\
1138 & Health and social work (85) & 0 & 0 & 0 & Other \\
1139 & Incineration of waste: Food & 0 & 1 & 0 & Manufacturing \\
1140 & Incineration of waste: Paper & 0 & 1 & 0 & Manufacturing \\
1141 & Incineration of waste: Plastic & 0 & 1 & 0 & Manufacturing \\
1142 & Incineration of waste: Metals and Inert materials & 0 & 1 & 0 & Manufacturing \\
1143 & Incineration of waste: Textiles & 0 & 1 & 0 & Manufacturing \\
1144 & Incineration of waste: Wood & 0 & 1 & 0 & Manufacturing \\
1145 & Incineration of waste: Oil/Hazardous waste & 0 & 1 & 0 & Manufacturing \\
1146 & Biogasification of food waste, incl. land application & 0 & 0 & 1 & Waste \\
1147 & Biogasification of paper, incl. land application & 0 & 0 & 1 & Waste \\
1148 & Biogasification of sewage slugde, incl. land application & 0 & 0 & 1 & Waste \\
1149 & Composting of food waste, incl. land application & 0 & 0 & 1 & Waste \\
1150 & Composting of paper and wood, incl. land application & 0 & 0 & 1 & Waste \\
1151 & Waste water treatment, food & 0 & 0 & 1 & Waste \\
1152 & Waste water treatment, other & 0 & 0 & 1 & Waste \\
1153 & Landfill of waste: Food & 0 & 0 & 1 & Waste \\
1154 & Landfill of waste: Paper & 0 & 0 & 1 & Waste \\
1155 & Landfill of waste: Plastic & 0 & 0 & 1 & Waste \\
1156 & Landfill of waste: Inert/metal/hazardous & 0 & 0 & 1 & Waste \\
1157 & Landfill of waste: Textiles & 0 & 0 & 1 & Waste \\
1158 & Landfill of waste: Wood & 0 & 0 & 1 & Waste \\
1159 & Activities of membership organisation n.e.c. (91) & 0 & 0 & 0 & Other \\
1160 & Recreational, cultural and sporting activities (92) & 0 & 0 & 0 & Other \\
1161 & Other service activities (93) & 0 & 0 & 0 & Other \\
1162 & Private households with employed persons (95) & 0 & 0 & 0 & Other \\
1163 & Extra-territorial organizations and bodies & 0 & 0 & 0 & Other \\
\end{longtable}
}

The table also reports the green \textit{policy-sector} classification used throughout the analysis. This classification provides a broad and policy-relevant aggregation of industries, reflecting the categories commonly employed in international climate agreements, policy documents, and emissions-reduction targets. Climate policy objectives are typically formulated in terms of broad emission-generating sectors rather than individual industries. Accordingly, we distinguish the following sector groups: Agriculture and Forestry, Electricity and Heat, Fossil Inputs, Manufacturing, Mining, Transport, and Waste. An additional category, labelled ``Other'', includes sectors that are generally not the primary focus of emissions-reduction policies, such as most service activities. More precisely, in the post-2025 scenarios, ``economy-wide'' coverage shall refer to all industries assigned to one of the policy-sector categories reported in the industry-classification table, excluding industries classified as ``Other''. For the European Union, scenario coverage instead depends on the EU ETS and EU ESR indicators: an industry is covered if it belongs to either regime.

\section{Model calibration and technical details}
\label{sec:S_model_calibration}\label{SM-sec:S_model_calibration}

\subsection*{Benchmark equilibrium and expenditure shares}

This section describes the calibration of the nested CES production and utility functions at the benchmark equilibrium and the allocation of inputs and consumption out of equilibrium.

The notation follows the model definition: $i \in \F$ denotes a producing firm, $j \in \M$ denotes an input good, $h \in \H$ denotes a household, and $s$ denotes a nest in the corresponding partition of goods.

The calibration is performed on the benchmark input--output equilibrium. Let $p_j$ denote the benchmark price of good $j$ and let $x_{ji}$ denote the quantity of good $j$ used by firm $i$. Total nominal expenditure by firm $i$ is
\begin{equation}
    E_i = \sum_{j \in \M} p_j x_{ji}.
\end{equation}
For a nest $s \in S_i$, the nominal expenditure of firm $i$ on that nest is
\begin{equation}
    E_{s,i} = \sum_{j \in s} p_j x_{ji}.
\end{equation}
The corresponding expenditure shares are
\begin{equation}
    \omega_{j,i} = \frac{p_j x_{ji}}{E_i},
    \qquad
    \Omega_{s,i} = \frac{E_{s,i}}{E_i}.
\end{equation}
The same notation is used for households, replacing the firm index $i$ with the household index $h$ and using observed benchmark consumption $x_{jh}$.

\subsection*{Calibration of nested CES production and utility functions}

For each firm $i \in \F$, production is represented by the nested CES technology
\begin{equation}
    f_i(x_i) = \Lambda_i \phi_i(x_i),
\end{equation}
with
\begin{equation}
    \phi_i(x_i)=
    \left[
      \sum_{s \in S_i}
      \left(
        \sum_{j \in s} \beta_{ji} x_{ji}^{\rho_s}
      \right)^{\theta / \rho_s}
    \right]^{1/\theta}.
\end{equation}
At the benchmark equilibrium, productivity is normalised to $\Lambda_i=1$. The parameters $\beta_{ji}$ are calibrated so that the observed input vector reproduces benchmark output $y_i$ and benchmark expenditure shares.

Therefore, after solving the first-order conditions, the $\beta$-parameters for the production function of firm $i$ are
\begin{equation}
    \label{eq:beta-production-calibration}
    \beta_{ji} = 
    \frac{p_{j}x_{ji}}{\sum_k p_{k}x_{ki}} 
    \left [ \frac{\sum_{k \in s} p_{k}x_{k,i}}{\sum_k p_{k}x_{k,i}} \right]^{\frac{\rho_s-\theta}{\theta}}
    \left (\frac{y_i}{x_{ji}}\right)^{\rho_s} =
    \omega_{j,i}
    \Omega_{s,i}^{\frac{\rho_s-\theta}{\theta}}
    \left(\frac{y_i}{x_{ji}}\right)^{\rho_s}\;,
\end{equation}
where $j$ denotes an input belonging to nest $s$, and $s \in S_i$ indexes the set of production nests of firm $i$. The calibration procedure to maximise household utility follows the same logic. 

The elasticity associated with the top-level aggregator is
\begin{equation}
    \sigma = \frac{1}{1-\theta} \;,
\end{equation}
while the elasticity inside nest $s$ is
\begin{equation}
    \sigma_s = \frac{1}{1-\rho_s}.
\end{equation}
These elasticities are defined whenever $\theta \neq 1$ and $\rho_s \neq 1$.

The calibrated matrices $\beta_{ji}$ and $\beta_{jh}$ are computed only for observed links in the benchmark input--output table. Non-observed links retain zero weight. Labour and capital services enter the production function through the value-added nest, consistent with the definition of $S_i$.

\subsection*{Weight parametrisation}

The above optimisation problem corresponds to the benchmark production problem: given prices and a nominal input budget, the firm chooses the input bundle that maximises production. We now consider the corresponding expenditure-allocation problem after prices change, taking the calibrated parameters $\beta_{ji}$, $\rho_s$, and $\theta$ as given.

For firm $i$, define the nest-specific aggregator
\begin{equation}
    G_{s,i}(x_i)=
    \left(
        \sum_{j \in s}\beta_{ji}x_{ji}^{\rho_s}
    \right)^{1/\rho_s},
    \qquad s \in S_i.
\end{equation}

The production aggregator is then
\begin{equation}
    \phi_i(x_i)=
    \left(
        \sum_{s \in S_i}G_{s,i}(x_i)^\theta
    \right)^{1/\theta}.
\end{equation}

Once the production function has been calibrated, the parameters $\beta_{ji}$, $\rho_s$, and $\theta$ remain fixed throughout the simulations. Out of equilibrium, firms do not re-optimise these parameters. Instead, they adjust their input bundles in response to changes in relative prices while preserving the calibrated production technology.

To characterise these adjustments, it is convenient to reformulate the firm's problem as a cost-minimisation problem. Given a target level of production, firms choose the least-cost combination of inputs subject to the nested CES production structure. The solution yields the optimal expenditure shares associated with each production nest and, subsequently, the optimal allocation of expenditure across individual inputs within each nest.

\subsection*{Cost minimisation and out-of-equilibrium input allocation}

For a given target output level normalised to one, the firm solves
\begin{equation}
\min_{\{x_{ji}\}}
C_i=\sum_{j\in\M} p_j x_{ji}
\qquad
\text{s.t.}
\qquad
\phi_i(x_i)=1.
\end{equation}

The optimisation can be solved recursively. First, expenditure is allocated across inputs within each nest. For a given level of the nest-specific aggregate $\bar G_{s,i}$, the corresponding Lagrangian is
\begin{equation}
\mathcal{L}_{s,i}
=
\sum_{j\in s} p_j x_{ji}
+
\lambda_s
\left(
\bar G_{s,i}-G_{s,i}
\right).
\end{equation}

The solution yields the nest-specific price index
\begin{equation}
\Pi_{s,i}
=
\left(
\sum_{j\in s}
p_j^{\,1-\sigma_s}
\beta_{ji}^{\,\sigma_s}
\right)^{\frac{1}{1-\sigma_s}},
\end{equation}
where
\begin{equation}
\sigma_s=\frac{1}{1-\rho_s}.
\end{equation}
The nest-specific price indices $\Pi_{s,i}$ represent the minimum unit cost of obtaining one unit of the composite input associated with nest $s$.

In a subsequent step, expenditure is allocated across nests. The corresponding optimisation problem is
\begin{equation}
\mathcal{L}_i
=
\sum_{s\in S_i}
\Pi_{s,i}G_{s,i}
+
\lambda
\left(
1-\phi_i
\right).
\end{equation}

Solving this problem yields the unit-cost function
\begin{equation}
C_i
=
\left[
\sum_{s\in S_i}
\Pi_{s,i}^{\,1-\sigma}
\right]^{\frac{1}{1-\sigma}}
=
\left[
\sum_{s\in S_i}
\left(
\sum_{j\in s}
p_j^{\,1-\sigma_s}
\beta_{ji}^{\,\sigma_s}
\right)^{\frac{1-\sigma}{1-\sigma_s}}
\right]^{\frac{1}{1-\sigma}} \; .
\end{equation}
The unit-cost function summarises the minimum expenditure required to produce one unit of output given the prevailing vector of input prices. Applying Shephard's lemma yields the optimal conditional demand for each input,
\begin{equation}
x_{ji}^{*}
=
\frac{\partial C_i}{\partial p_j},
\end{equation}
which determines the expenditure shares and input coefficients used in the out-of-equilibrium dynamics.

\section{Production-network measures}
\label{sec:S_network_measures}\label{SM-sec:S_network_measures}

To analyse the structural changes in the input--output network, we measure upstreamness and downstreamness, which capture the relative position of sectors along the production chain, as well as network centrality measures such as in- and out-strength and Katz centrality (forward and backward). These network statistics help to quantify the intensity, reach, and systemic importance of sectors within the network.

Under the standard assumption that intermediate input requirements are proportional to sectoral output, let
\[
\mathbf{Z}_t=(z_{ij,t})_{i,j\in\mathcal F}
\]
denote the matrix of real input coefficients, where $z_{ij}^t=x_{ij}^t/y_j^t$ measures the amount of input supplied by firm $i$ per unit of real output produced by firm $j$. These coefficients are endogenous and can vary over time as firms adjust their input combinations in response to changes in relative prices. The matrix $\mathbf{Z}_t \in \mathbb{R}^{n\times n}$ therefore describes the firm-to-firm component of the production network.

The input--output accounting identity of the production system can then be written as
\begin{equation}
    Y_t=\mathbf{Z}_tY_t+F_t.
\end{equation}
In this expression, $Y_t$ denotes the vector of gross output and $F_t$ the vector of final demand, both expressed in real terms. The vector $\mathbf{Z}_tY_t$ therefore corresponds to intermediate demand generated by production activities.

We define the production network as the matrix $\mathbf{Z}$, omitting the time subscript whenever no confusion arises. The element $z_{ij}$ measures the intensity of the production linkage from supplier $i$ to customer $j$, and the resulting weighted directed network describes the structure of inter-firm production relationships.

\subsection*{Strength and Katz centrality}
The out-strength and in-strength of industry $i$ are defined as the sums of its outgoing and incoming links:
\begin{align}    
\mathbf{s}^{out} &= \mathbf{Z} \mathbf{1} \\
\mathbf{s}^{in} &= \mathbf{Z}^{\top} \mathbf{1} \;.
\end{align}
By construction of the matrix $\mathbf{Z}$, in-strength is bounded between 0 and 1 and measures the share of intermediate inputs in total production. Industries with high in-strength rely more heavily on intermediate inputs than on value added, and therefore exhibit stronger direct production dependencies on other sectors. In contrast, out-strength is not bounded by unity. It measures the extent to which an industry's output is used as an intermediate input by other industries, relative to those industries' total production. High out-strength identifies sectors that play an important role as direct suppliers within the production network and whose outputs are widely embedded in other industries' production processes.

Forward Katz centrality captures the importance of an industry as a supplier through both direct and indirect production linkages. Industries with high centrality values are connected to sectors that themselves supply many other sectors, implying that their output propagates widely throughout the economy. This is defined as
\begin{equation}
    \mathbf{k}^{forward} = (I - \alpha \mathbf{Z})^{-1} \mathbf{1} \;;
\end{equation}
This expression converges when $\alpha < \frac{1}{\rho(\mathbf{Z})}$, where $\rho(\mathbf{Z})$ denotes the spectral radius of the production-network matrix. 

Backward Katz centrality measures the extent to which an industry's production depends on a broad network of upstream suppliers. High values indicate sectors whose production processes are embedded within long and interconnected supply chains. This is defined as
\begin{equation}
    \mathbf{k}^{backward} = (I - \alpha \mathbf{Z}^{\top})^{-1} \mathbf{1} \;.
\end{equation}

Because we compare network statistics across scenarios, we need a common attenuation parameter $\alpha$, which is defined as:
\begin{equation}
\alpha=\frac{0.5}{\max_{r\in\mathcal R}\rho(\mathbf Z_r)},
\end{equation}
where $\rho(\mathbf Z_r)$ is the spectral radius of network $r$. This choice satisfies the convergence condition for every network in $\mathcal R$. The same $\alpha$ applies to forward and backward centrality because $\mathbf Z_t$ and $\mathbf Z_t^{\top}$ have the same spectrum. Katz vectors are not divided by the maximum centrality of their respective networks. This preserves comparisons across BAU and HWTP without introducing a network-specific rescaling.

\subsection*{Upstreamness and downstreamness}

Following \citet{antras2012measuring}, we also measure \textit{Upstreamness} and \textit{Downstreamness}, two indicators that provide more information about the position of industries within the production network. Unlike strength and centrality measures, which quantify the intensity and systemic importance of production linkages, upstreamness and downstreamness describe where industries are located along the production process, from primary inputs to final demand.

Upstreamness measures the average distance of a sector from final demand and can be interpreted as the expected number of production stages remaining before a sector's output reaches final consumers. Industries with high upstreamness are located further away from final demand and typically supply intermediate inputs that undergo multiple transformations before reaching consumers. Examples include mining, basic materials, and energy-producing industries. This is defined as
\begin{equation}\label{upstream}
U = [\mathbf{I} - \mathbf{Y}^{-1}\mathbf{Z}\mathbf{Y}]^{-1}\mathbf{1} \;.
\end{equation}

Downstreamness captures the average distance from primary factors of production, such as labour, capital, and natural resources. It therefore measures how far an industry is located from the origin of production chains. Industries with high downstreamness tend to operate closer to final demand and combine inputs originating from multiple upstream sectors. Typical examples include consumer-oriented manufacturing and service activities. This is defined as
\begin{equation}\label{downstream}
D = [\mathbf{I} - \mathbf{Y}^{-1}\mathbf{Z}^{\top}\mathbf{Y}]^{-1}\mathbf{1} \;.
\end{equation}
In Equations~(\ref{upstream}) and (\ref{downstream}), $\mathbf I$ denotes the identity matrix, $\mathbf Y=\mathrm{diag}(Y)$ is the diagonal matrix of sectoral gross output, $\mathbf Y^{-1}$ is the corresponding inverse diagonal matrix, and $\mathbf 1$ is a column vector of ones.

\section{Climate-policy implementation and calibration}
\label{sec:S_policy}\label{SM-sec:S_policy}

\subsection*{Historical policy calibration, 2019--2025}

For 2019--2025, the simulations implement the observed evolution of carbon taxes and emissions-allowance prices. Where policy is specified through an emissions-reduction target, the corresponding cap is imposed exogenously; the carbon price required to meet that cap is generated endogenously by the model.

\subsection*{Post-2025 regional policy scenarios and targets}

Table~\ref{tab:S_policy_scenarios} summarises the climate-policy instruments used for the historical calibration and the assumptions adopted for the post-2025 scenarios. The policy trajectories end in 2035; the resulting policy settings are then held fixed until the end of the simulation in 2040.

The information reported in Table~\ref{tab:S_policy_scenarios} is primarily based on the UNFCCC NDC Registry, which documents countries' climate commitments under the Paris Agreement. We complement this information with additional datasets on carbon-pricing policies and emissions trading systems, including \citet{hattori2022iges} and \citet{dafnomilis2024climate}. 

\begin{table}[h!]
\centering
\caption{\footnotesize Regional climate-policy scenarios: policy instruments, coverage, timing, and targets.}
\label{tab:S_policy_scenarios}\label{SM-tab:S_policy_scenarios}
\resizebox{\textwidth}{!}{%
\begin{tabular}{llllclclclc}
\toprule
Market & Scenario policy scope & Calibration instrument & Start year & Calibration target & Calibration unit & Calibration year & Scenario instrument & Scenario target & Scenario unit & Final year \\
\toprule
\multicolumn{11}{c}{\textit{BAU: Business as usual}} \\
EU27 & EU-ETS & ETS & 2019 & -33 & GHGE-cut(\%) & 2025 & ETS & -- & GHGE-cut(\%) & 2035 \\
EU27 & EU-ESR & ETS & 2019 & -20 & GHGE-cut(\%) & 2025 & ETS & -- & GHGE-cut(\%) & 2035 \\
GB & Economy wide & ETS & 2019 & -33 & GHGE-cut(\%) & 2025 & ETS & -- & GHGE-cut(\%) & 2035 \\
CN & Electricity \& Heat & TAX & 2020 & 5 & EUR/tCO2 & 2025 & TAX & -- & EUR/tCO2 & 2035 \\
BR & Manufacturing & ETS & 2023 & -6 & GHGE-cut(\%) & 2025 & ETS & -- & GHGE-cut(\%) & 2035 \\
CA & Manufacturing & TAX & 2022 & 35 & EUR/tCO2 & 2025 & TAX & -- & EUR/tCO2 & 2035 \\
ID & Economy wide & TAX & 2021 & 2 & EUR/tCO2 & 2025 & TAX & -- & EUR/tCO2 & 2035 \\
JP & Electricity \& Heat & TAX & 2019 & 5 & EUR/tCO2 & 2025 & TAX & -- & EUR/tCO2 & 2035 \\
KR & Economy wide & ETS & 2019 & -6 & GHGE-cut(\%) & 2025 & ETS & -- & GHGE-cut(\%) & 2035 \\
MX & Electricity \& Heat & TAX & 2019 & 3 & EUR/tCO2 & 2025 & TAX & -- & EUR/tCO2 & 2035 \\
ZA & Economy wide & TAX & 2024 & 10 & EUR/tCO2 & 2025 & TAX & -- & EUR/tCO2 & 2035 \\
\toprule
\multicolumn{11}{c}{\textit{NDC: Nationally Determined Contributions}} \\
EU27 & EU-ETS & ETS & 2019 & -33 & GHGE-cut(\%) & 2025 & ETS & -59 & GHGE-cut(\%) & 2035 \\
EU27 & EU-ESR & ETS & 2019 & -20 & GHGE-cut(\%) & 2025 & ETS & -59 & GHGE-cut(\%) & 2035 \\
GB & Economy wide & ETS & 2019 & -33 & GHGE-cut(\%) & 2025 & ETS & -67 & GHGE-cut(\%) & 2035 \\
CN & Economy wide & TAX & 2020 & 5 & EUR/tCO2 & 2025 & ETS & -52 & GHGE-cut(\%) & 2035 \\
BR & Economy wide & ETS & 2023 & -6 & GHGE-cut(\%) & 2025 & ETS & -68 & GHGE-cut(\%) & 2035 \\
CA & Economy wide & TAX & 2022 & 35 & EUR/tCO2 & 2025 & ETS & -52 & GHGE-cut(\%) & 2035 \\
ID & Economy wide & TAX & 2021 & 2 & EUR/tCO2 & 2025 & ETS & -40 & GHGE-cut(\%) & 2035 \\
JP & Economy wide & TAX & 2019 & 5 & EUR/tCO2 & 2025 & ETS & -53 & GHGE-cut(\%) & 2035 \\
KR & Economy wide & ETS & 2019 & -6 & GHGE-cut(\%) & 2025 & ETS & -53 & GHGE-cut(\%) & 2035 \\
MX & Economy wide & TAX & 2019 & 3 & EUR/tCO2 & 2025 & ETS & -30 & GHGE-cut(\%) & 2035 \\
ZA & Economy wide & TAX & 2024 & 10 & EUR/tCO2 & 2025 & ETS & -27 & GHGE-cut(\%) & 2035 \\
AU & Economy wide & ETS & 2026 & 0 & GHGE-cut(\%) & 2025 & ETS & -67 & GHGE-cut(\%) & 2035 \\
IN & Economy wide & ETS & 2026 & 0 & GHGE-cut(\%) & 2025 & ETS & -50 & GHGE-cut(\%) & 2035 \\
RU & Economy wide & ETS & 2026 & 0 & GHGE-cut(\%) & 2025 & ETS & -55 & GHGE-cut(\%) & 2035 \\
TR & Economy wide & ETS & 2026 & 0 & GHGE-cut(\%) & 2025 & ETS & -30 & GHGE-cut(\%) & 2035 \\
TW & Economy wide & ETS & 2026 & 0 & GHGE-cut(\%) & 2025 & ETS & -40 & GHGE-cut(\%) & 2035 \\
US & Economy wide & ETS & 2026 & 0 & GHGE-cut(\%) & 2025 & ETS & -53 & GHGE-cut(\%) & 2035 \\
WA & Economy wide & ETS & 2026 & 0 & GHGE-cut(\%) & 2025 & ETS & -40 & GHGE-cut(\%) & 2035 \\
WE & Economy wide & ETS & 2026 & 0 & GHGE-cut(\%) & 2025 & ETS & -40 & GHGE-cut(\%) & 2035 \\
WF & Economy wide & ETS & 2026 & 0 & GHGE-cut(\%) & 2025 & ETS & -40 & GHGE-cut(\%) & 2035 \\
WL & Economy wide & ETS & 2026 & 0 & GHGE-cut(\%) & 2025 & ETS & -40 & GHGE-cut(\%) & 2035 \\
WM & Economy wide & ETS & 2026 & 0 & GHGE-cut(\%) & 2025 & ETS & -40 & GHGE-cut(\%) & 2035 \\
\toprule
\multicolumn{11}{c}{\textit{HWTP: Highway to Paris}} \\
EU27 & EU-ETS & ETS & 2019 & -33 & GHGE-cut(\%) & 2025 & TAX & 200 & EUR/tCO2 & 2035 \\
EU27 & EU-ESR & ETS & 2019 & -20 & GHGE-cut(\%) & 2025 & TAX & 200 & EUR/tCO2 & 2035 \\
GB & Economy wide & ETS & 2019 & -33 & GHGE-cut(\%) & 2025 & TAX & 200 & EUR/tCO2 & 2035 \\
CN & Economy wide & TAX & 2020 & 5 & EUR/tCO2 & 2025 & TAX & 200 & EUR/tCO2 & 2035 \\
BR & Economy wide & ETS & 2023 & -6 & GHGE-cut(\%) & 2025 & TAX & 200 & EUR/tCO2 & 2035 \\
CA & Economy wide & TAX & 2022 & 35 & EUR/tCO2 & 2025 & TAX & 200 & EUR/tCO2 & 2035 \\
ID & Economy wide & TAX & 2021 & 2 & EUR/tCO2 & 2025 & TAX & 200 & EUR/tCO2 & 2035 \\
JP & Economy wide & TAX & 2019 & 5 & EUR/tCO2 & 2025 & TAX & 200 & EUR/tCO2 & 2035 \\
KR & Economy wide & ETS & 2019 & -6 & GHGE-cut(\%) & 2025 & TAX & 200 & EUR/tCO2 & 2035 \\
MX & Economy wide & TAX & 2019 & 3 & EUR/tCO2 & 2025 & TAX & 200 & EUR/tCO2 & 2035 \\
ZA & Economy wide & TAX & 2024 & 10 & EUR/tCO2 & 2025 & TAX & 200 & EUR/tCO2 & 2035 \\
AU & Economy wide & ETS & 2026 & 0 & GHGE-cut(\%) & 2025 & TAX & 200 & EUR/tCO2 & 2035 \\
IN & Economy wide & ETS & 2026 & 0 & GHGE-cut(\%) & 2025 & TAX & 200 & EUR/tCO2 & 2035 \\
RU & Economy wide & ETS & 2026 & 0 & GHGE-cut(\%) & 2025 & TAX & 200 & EUR/tCO2 & 2035 \\
TR & Economy wide & ETS & 2026 & 0 & GHGE-cut(\%) & 2025 & TAX & 200 & EUR/tCO2 & 2035 \\
TW & Economy wide & ETS & 2026 & 0 & GHGE-cut(\%) & 2025 & TAX & 200 & EUR/tCO2 & 2035 \\
US & Economy wide & ETS & 2026 & 0 & GHGE-cut(\%) & 2025 & TAX & 200 & EUR/tCO2 & 2035 \\
WA & Economy wide & ETS & 2026 & 0 & GHGE-cut(\%) & 2025 & TAX & 200 & EUR/tCO2 & 2035 \\
WE & Economy wide & ETS & 2026 & 0 & GHGE-cut(\%) & 2025 & TAX & 200 & EUR/tCO2 & 2035 \\
WF & Economy wide & ETS & 2026 & 0 & GHGE-cut(\%) & 2025 & TAX & 200 & EUR/tCO2 & 2035 \\
WL & Economy wide & ETS & 2026 & 0 & GHGE-cut(\%) & 2025 & TAX & 200 & EUR/tCO2 & 2035 \\
WM & Economy wide & ETS & 2026 & 0 & GHGE-cut(\%) & 2025 & TAX & 200 & EUR/tCO2 & 2035 \\
\bottomrule
\end{tabular}%
}
\end{table}

Table~\ref{tab:S_policy_scenarios} documents the policy instruments, sectoral coverage, implementation dates, and quantitative targets used in the three climate-policy scenarios. All scenarios share the same calibrated policy path through 2025. Thereafter, BAU holds the 2025 policy settings constant, NDC implements country-specific commitments primarily as emissions caps, and HWTP introduces a harmonised carbon-tax path that reaches EUR~200/tCO$_2$ in 2035. NDC and HWTP are alternative policy configurations with different instruments and coverage, rather than a simple ranking of policy stringency.

\subsection*{Calibration diagnostics}

\begin{figure}[h!]
\centering
\includegraphics[width=0.99\linewidth]{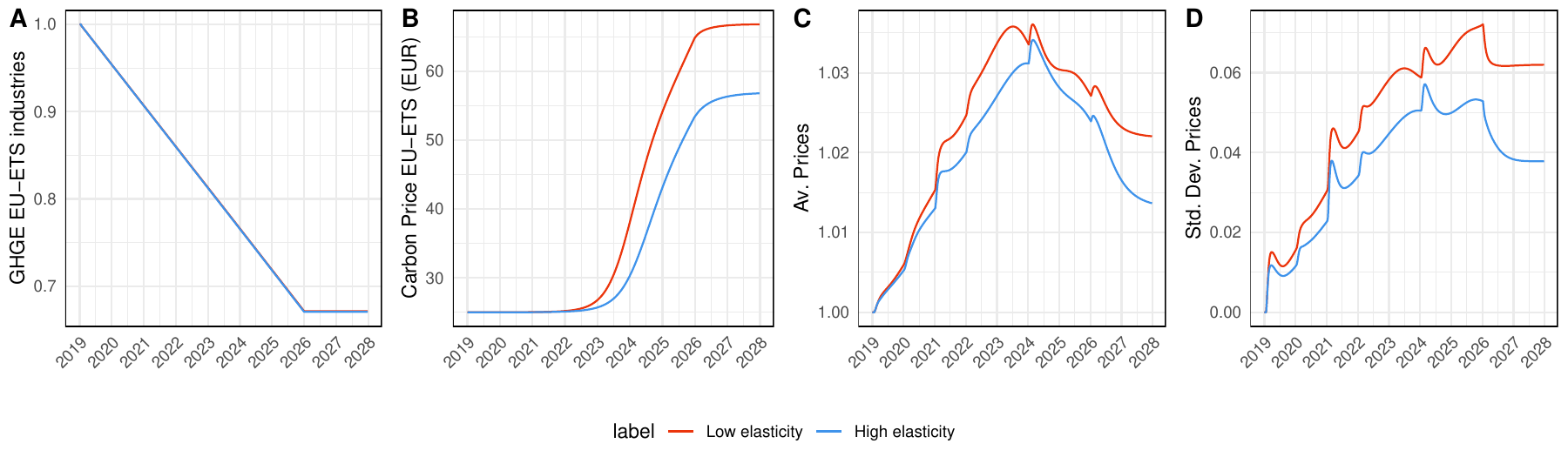}
\caption{EU ETS calibration under alternative elasticity assumptions, 2019--2025. Panel A reports emissions subject to the common exogenous cap; Panel B reports the endogenous carbon-price response. Panels C and D report the mean and cross-sector standard deviation of output prices.}
\label{fig:S_calibration}
\end{figure}

Figure~\ref{fig:S_calibration} illustrates the dynamic adjustment of emissions and carbon prices within the EU ETS under alternative elasticity calibrations. Panel~A shows greenhouse-gas emissions from EU ETS industries under the low- and high-elasticity calibrations. The emissions cap is imposed exogenously and tightened to deliver a 33\% reduction relative to 2019 by 2025, so the emissions paths are nearly identical across the two calibrations. Panel~B reports the carbon prices generated endogenously to satisfy that cap. The low-elasticity calibration yields systematically higher prices because substitution possibilities are more limited. The simulated 2025 carbon prices lie between approximately EUR~55 and EUR~70/tCO$_2$eq. For comparison, observed EU ETS allowance prices ranged from EUR~63 at the end of March 2025 to EUR~87 at the end of December 2025 and have fluctuated around EUR~70 since 2022.

Panels~C and~D of Figure~\ref{fig:S_calibration} present the evolution of average prices and their dispersion, respectively. Once policy adjustments cease in the final iterations corresponding to 2025, with caps and carbon taxes held fixed, output prices gradually converge towards a new equilibrium.

\section{Additional simulation results}
\label{sec:S_results}\label{SM-sec:S_results}

\subsection*{Global macroeconomic dynamics}

\begin{figure}[h!]
\centering
\includegraphics[width=0.99\linewidth]{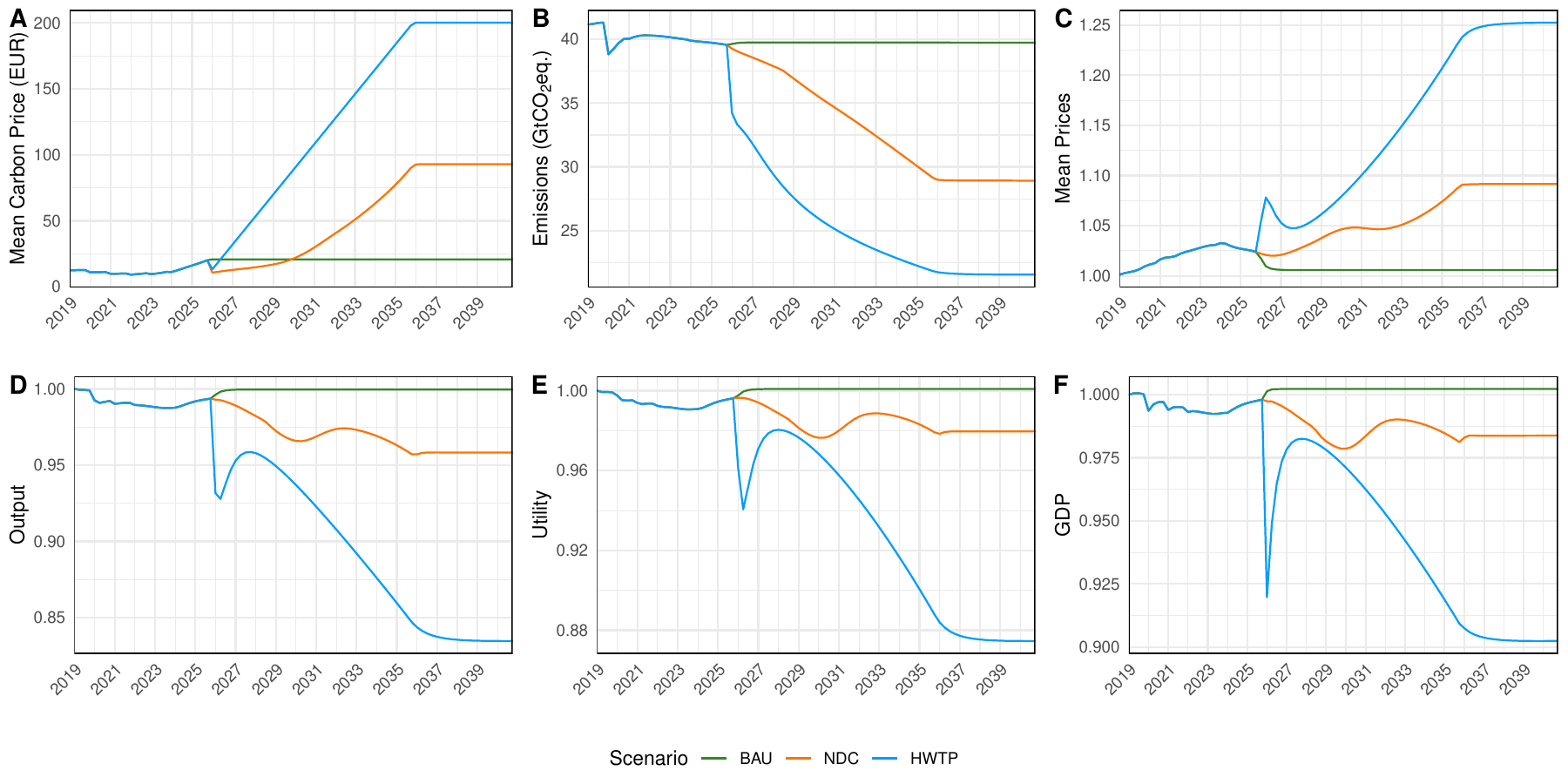}
\caption{Global macroeconomic dynamics under the BAU, NDC, and HWTP scenarios, 2019--2040. Panels A--F report, respectively, the mean carbon price across operating carbon markets (EUR/tCO$_2$eq), global greenhouse-gas emissions (GtCO$_2$eq), and indices of mean prices, output, household utility, and GDP. The four indices are normalised to one in 2019.}
\label{fig:S_macro}\label{SM-fig:S_macro}
\end{figure}

This section reports global aggregate dynamics that complement the sectoral results in the main manuscript. Figure~\ref{fig:S_macro} shows carbon prices, greenhouse-gas emissions, prices, output, household utility, and GDP under BAU, NDC, and HWTP.

Panel~A reports the mean carbon price across operating carbon markets. The three scenarios share the same historical calibration through 2025. From 2026, the BAU price remains close to its 2025 level, the NDC path reaches approximately EUR~90/tCO$_2$eq in 2035, and the HWTP path reaches EUR~200/tCO$_2$eq in 2035. These values then remain stable until the end of the simulation in 2040.\footnote{The small discontinuities are composition effects caused by the progressive inclusion of carbon markets with lower initial prices, rather than relaxations of climate policy.}

Panel~B reports global greenhouse-gas emissions. BAU emissions stabilise at just under 40~GtCO$_2$eq. By the end of the policy trajectory in 2035, emissions fall to approximately 29~GtCO$_2$eq under NDC and to approximately 22~GtCO$_2$eq under HWTP, after which all three paths remain broadly stable through 2040. The different paths reflect the distinct instruments, geographical coverage, and sectoral coverage of the NDC and HWTP configurations.

Panels~C--F report the evolution of mean prices, output, household utility, and GDP. Relative to BAU at the end of the simulation, both NDC and HWTP raise mean prices and reduce output, utility, and GDP. The changes are visibly larger under HWTP than under NDC in each panel. The series also show short-run adjustments when the post-2025 policies are introduced, followed by gradual convergence once the policy trajectories end in 2035.

\subsection*{Sensitivity to substitution elasticities}

To assess how the aggregate results depend on substitution possibilities, we conduct a sensitivity exercise around the benchmark calibration, keeping all other parameters and scenario definitions unchanged. We consider low, benchmark, and high values of the top-level elasticity of substitution, $\sigma$, and of the elasticity within the energy nest, $\sigma_{\mathrm{energy}}$. These configurations span the sensitivity ranges reported in the main text and should be interpreted as boundary cases rather than as probabilistic confidence intervals.

\begin{figure}[h!]
\centering
\includegraphics[width=0.99\linewidth]{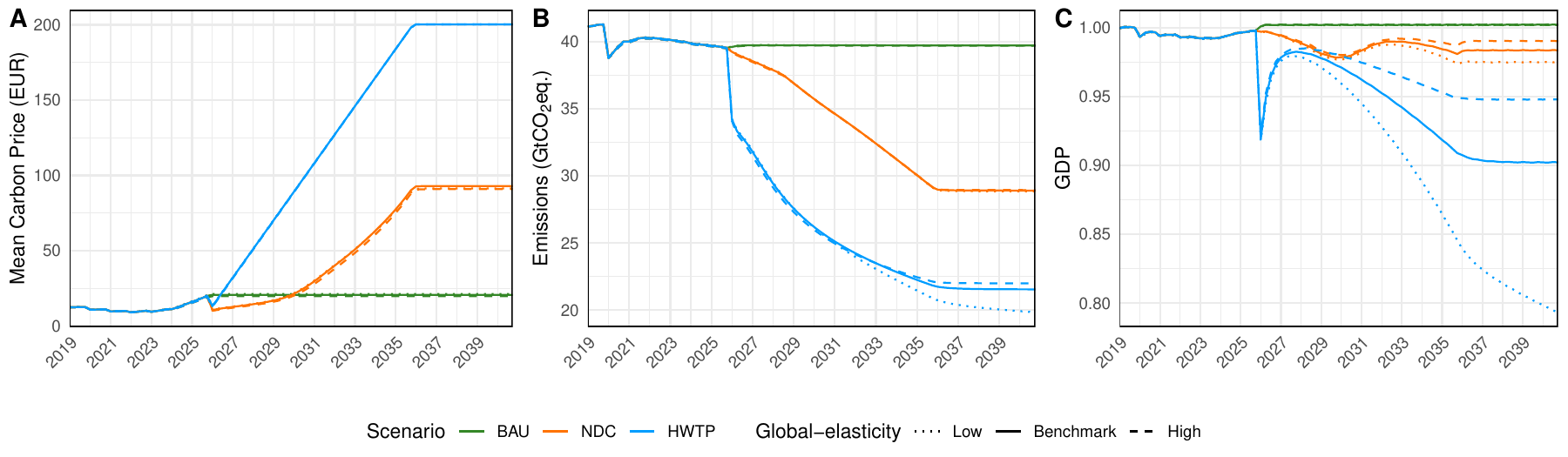}
\caption{Sensitivity of global aggregate dynamics to the top-level elasticity of substitution, 2019--2040. Panels A--C report, respectively, the mean carbon price across operating carbon markets, global greenhouse-gas emissions, and the global GDP index, normalised to one in 2019. Line types distinguish the low ($\sigma=0.25$), benchmark ($\sigma=0.33$), and high ($\sigma=0.50$) elasticity configurations. All other parameters are held at their benchmark values.}
\label{fig:global_elasticity_sensitivity}\label{SM-fig:global_elasticity_sensitivity}
\end{figure}

Figure~\ref{fig:global_elasticity_sensitivity} shows that alternative values of the global elasticity preserve the qualitative ordering of the scenarios but affect the magnitude of the adjustment. Under the NDC configuration, emissions are largely determined by the imposed emissions constraints, and sensitivity therefore appears primarily in the endogenous carbon price and GDP. Under HWTP, the carbon-price path is fixed by construction, so lower substitution possibilities produce larger reductions in emissions but also substantially greater GDP losses. The additional emissions reduction under low elasticity thus partly reflects a stronger contraction in economic activity.

\begin{figure}[h!]
\centering
\includegraphics[width=0.99\linewidth]{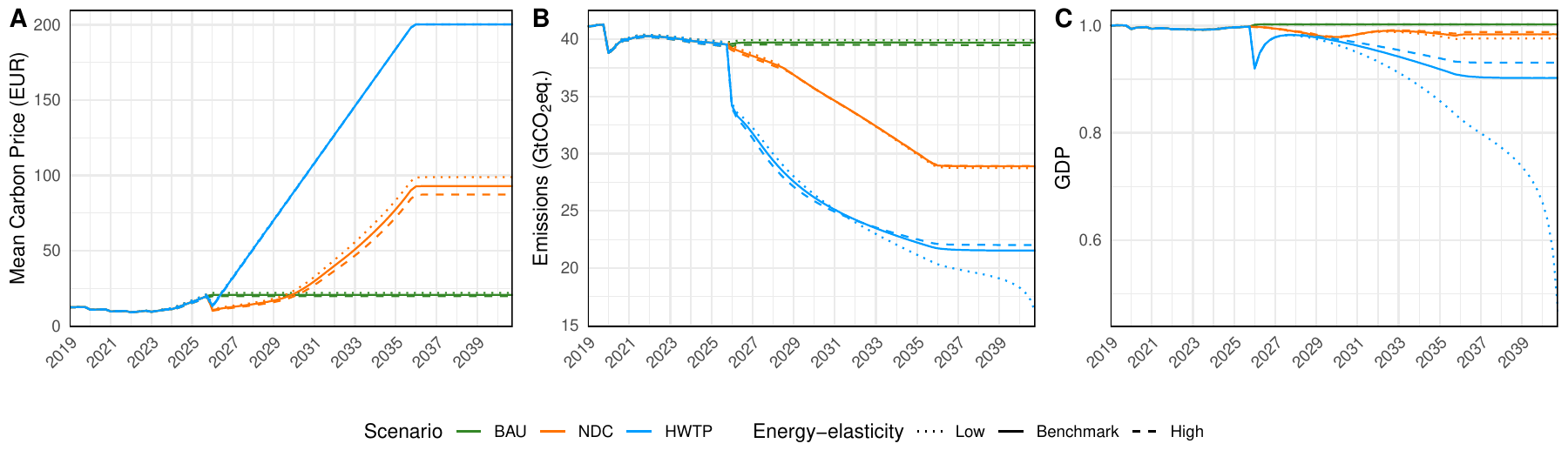}
\caption{Sensitivity of global aggregate dynamics to the elasticity of substitution within the energy nest, 2019--2040. Panels A--C report, respectively, the mean carbon price across operating carbon markets, global greenhouse-gas emissions, and the global GDP index, normalised to one in 2019. Line types distinguish the low ($\sigma_{\mathrm{energy}}=0.83$), benchmark ($\sigma_{\mathrm{energy}}=1.67$), and high ($\sigma_{\mathrm{energy}}=2.50$) elasticity configurations. All other parameters are held at their benchmark values.}
\label{fig:energy_elasticity_sensitivity}\label{SM-fig:energy_elasticity_sensitivity}
\end{figure}

Figure~\ref{fig:energy_elasticity_sensitivity} indicates large sensitivity to substitution possibilities within the energy nest. The benchmark and high-elasticity calibrations generate stable aggregate trajectories, whereas the low-elasticity configuration produces a strong and delayed adjustment under HWTP. When substitution between energy inputs is severely constrained, the production system has limited capacity to accommodate the imposed carbon-price trajectory, amplifying the contraction in GDP, showing the implication of energy-substitution possibilities for the economic adjustment to ambitious climate policies. It should therefore be interpreted as illustrating the limits of the production system's capacity to accommodate the imposed carbon-price path. 

\subsection*{Extended production-network results}

This section supplements the main production-network analysis with results for all country-sector observations and for broad non-energy sectors. All comparisons are between HWTP and BAU at the end of the simulation, $t_f$; they are not initial-period versus final-period comparisons. Because the feasible supplier--customer links are fixed in the model, the results describe production-network recomposition and sectoral repositioning within existing production links rather than unrestricted network rewiring.

\begin{figure}[h!]
\centering
\includegraphics[width=0.99\linewidth]{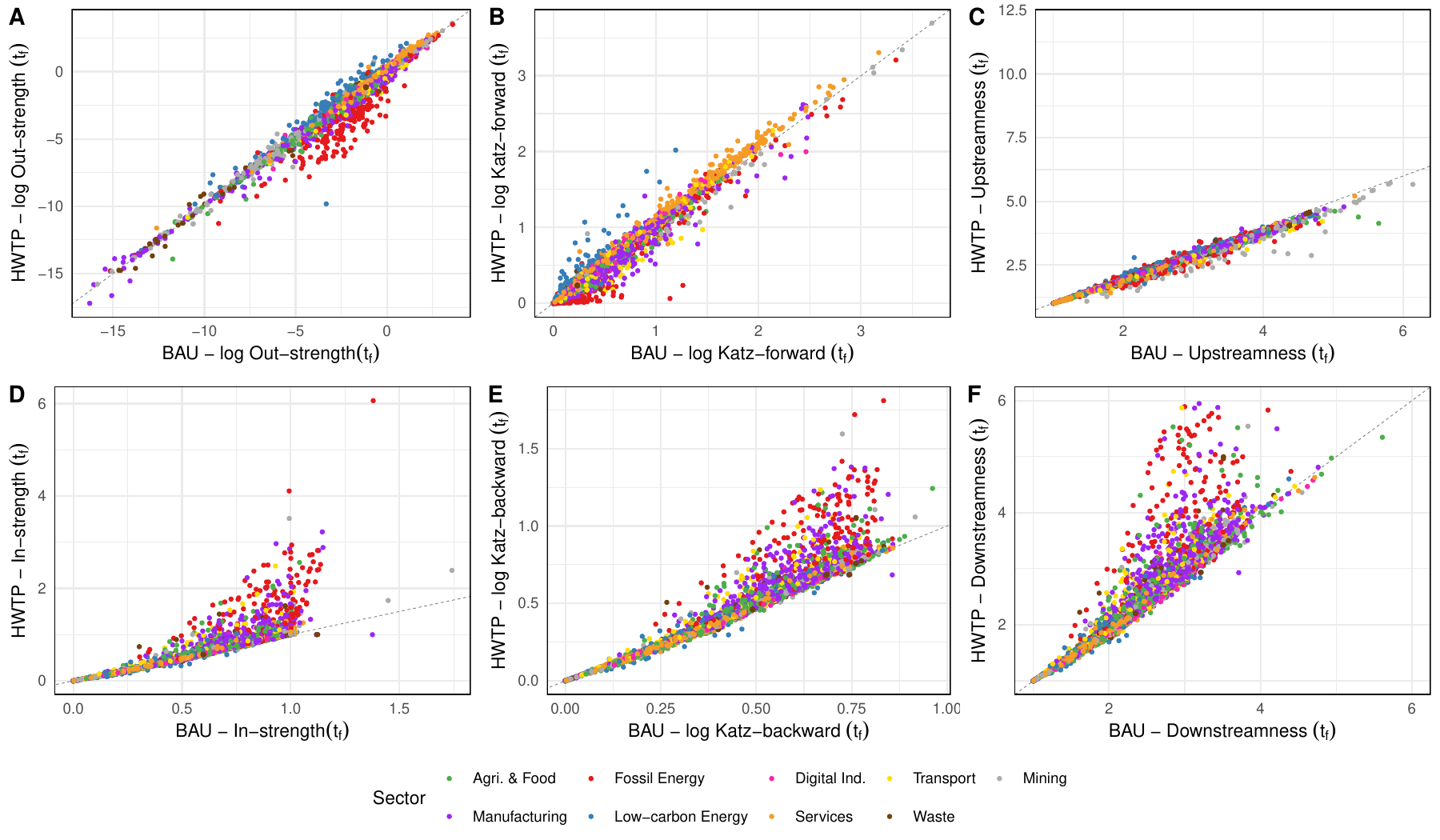}
\caption{Production-network statistics across all country-sector observations under HWTP versus BAU at $t_f$. The horizontal axis gives the BAU value and the vertical axis the HWTP value. Panels A, B, and E report the logarithms of out-strength, forward Katz centrality, and backward Katz centrality, respectively; Panels C, D, and F report upstreamness, in-strength, and downstreamness in levels. The dashed line is the 45-degree line.}
\label{fig:S_net_scatter}
\end{figure}

Figure~\ref{fig:S_net_scatter} plots each network statistic under HWTP against its BAU value at $t_f$. A large share of observations remains close to the 45-degree line, indicating persistence in the relative structure of the network, while deviations reveal heterogeneous changes in linkage intensity and production-chain position. Fossil-energy observations tend to lie below the line for log out-strength, log forward Katz centrality, and upstreamness, but above it for in-strength, log backward Katz centrality, and downstreamness. Low-carbon energy observations generally move in the opposite direction in the supplier dimension, with higher log out-strength and log forward centrality, while their other changes are smaller and more heterogeneous. These sectoral patterns coexist with substantial dispersion across countries and with observations on both sides of the line.

\begin{figure}[h!]
\centering
\includegraphics[width=0.99\linewidth]{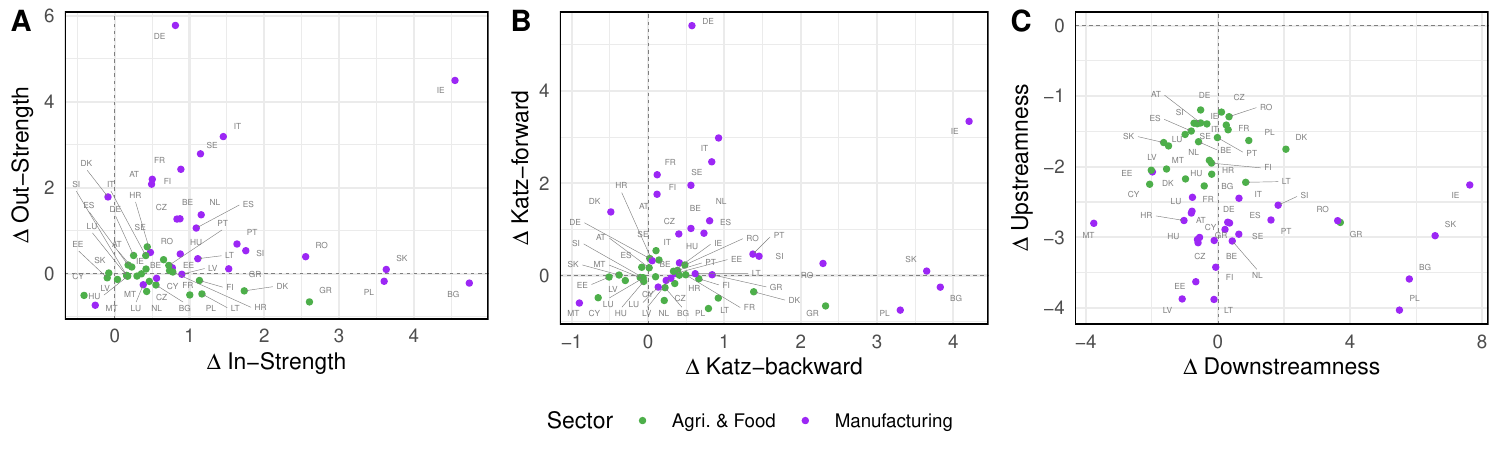}
\includegraphics[width=0.99\linewidth]{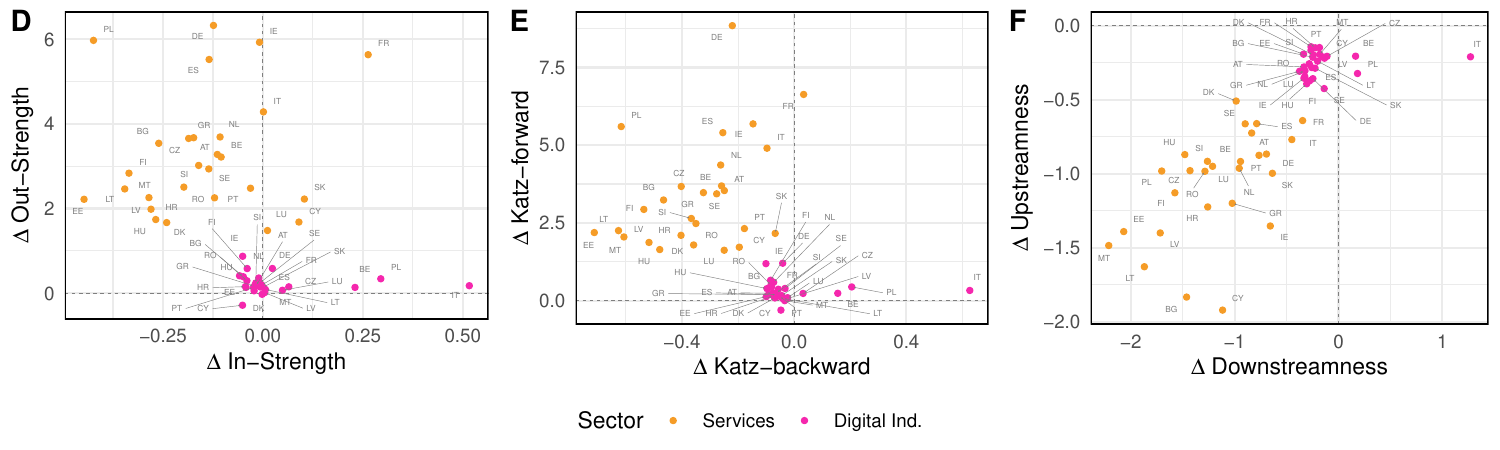}
\caption{Production-network adjustment under HWTP relative to BAU across EU countries at $t_f$. All axes report HWTP--BAU differences in levels. Panels A and D plot in-strength against out-strength; Panels B and E plot backward against forward Katz centrality; Panels C and F plot downstreamness against upstreamness. Panels A--C report agriculture and food and manufacturing; Panels D--F report services and digital industries.}
\label{fig:S_net_sectors}\label{SM-fig:S_net_sectors}
\end{figure}

Figure~\ref{fig:S_net_sectors} reports HWTP--BAU differences at $t_f$ for agriculture and food, manufacturing, services, and digital industries across EU countries. Manufacturing exhibits higher in-strength and out-strength in many countries, frequently accompanied by increases in both Katz measures. Its upstreamness generally declines, whereas downstreamness is markedly heterogeneous. Agriculture and food displays smaller and more heterogeneous changes in strength and Katz centrality, although upstreamness generally declines. Services show clear increases in out-strength and forward Katz centrality, while changes in in-strength and backward centrality are smaller, heterogeneous, and often negative; both upstreamness and downstreamness tend to decline, with cross-country variation. Digital industries remain comparatively close to zero in in-strength and backward centrality, show moderate and heterogeneous changes in the supplier dimension, and exhibit heterogeneous changes in downstreamness.

\clearpage
\bibliographystyle{chicago}
\bibliography{biblio_rev.bib}

\end{document}